\documentclass[a4paper,fleqn]{cas-dc}

\usepackage[authoryear]{natbib}
\usepackage[titletoc,title]{appendix}
\usepackage{float}
\usepackage{todonotes}

\def\tsc#1{\csdef{#1}{\textsc{\lowercase{#1}}\xspace}}
\tsc{WGM}
\tsc{QE}
\tsc{EP}
\tsc{PMS}
\tsc{BEC}
\tsc{DE}
\begin{document}
\let\WriteBookmarks\relax
\def\floatpagepagefraction{1}
\def\textpagefraction{.001}
\shorttitle{Identification and Measurement of Technical Debt Requirements in Software Development: A Systematic Literature Review}
\shortauthors{Ana Melo et~al.}

\title [mode = title]{Identification and Measurement of Technical Debt Requirements in Software Development: a Systematic Literature Review}                      
%\tnotemark[1,2]

%\tnotetext[1]{This document is the results of the research
 %  project funded by the National Science Foundation.}

%\tnotetext[2]{The second title footnote which is a longer text matter
 %  to fill through the whole text width and overflow into
 %  another line in the footnotes area of the first page.}

\author[1]{Ana Melo}[type=editor,
                        auid=000,bioid=1,
                       %prefix=Sir,
                        %role=Researcher,
                 orcid=0000-0003-4608-0651]
%\cormark[1]
\ead{accm@ecomp.poli.br}
%\ead[url]{www.cvr.cc, cvr@sayahna.org}

\credit{Conceptualization, Methodology, Data extraction, Data analysis, Writing - original draft \& editing}

\address[1]{University of Pernambuco, Recife, Brazil}

\author[1]{Roberta Fagundes}[%
  %role=Co-ordinator,
   %suffix=Jr,
   ]
\ead{roberta.fagundes@upe.br}
%\ead[URL]{www.sayahna.org}

\credit{Data analysis and Review}

\address[2]{LUT University, Lathi, Finland}

%\address[2]{Sayahna Foundation, Jagathy, Trivandrum 695014, India}

\author%
[2]
{Valentina Lenarduzzi}
%\cormark[2]
%\fnmark[1,3]
\ead{valentina.lenarduzzi@lut.fi}
%\ead[URL]{www.stmdocs.in}
\credit{Methodology, Review \& editing}

\author[1]{Wylliams Santos}[style=chinese]
\ead{wbs@upe.br}

\credit{Supervision, Data analysis, Writing - Review}

\cortext[cor1]{Corresponding author: Valentina Lenarduzzi}
%\cortext[cor2]{Principal corresponding author}
%\fntext[fn1]{This is the first author footnote. but is common to third
 % author as well.}
%\fntext[fn2]{Another author footnote, this is a very long footnote and
 % it should be a really long footnote. But this footnote is not yet
 % sufficiently long enough to make two lines of footnote text.}

%\nonumnote{This note has no numbers. In this work we demonstrate
 % the formation Y\_1 of a new type of polariton on the interface
 % between a cuprous oxide slab and a polystyrene micro-sphere placed
  %on the slab.
 % }

\begin{abstract}
\noindent \textbf{Context:} Technical Debt  requirements are related to the distance between the ideal value of the specification and the system's actual implementation, which are consequences of strategic decisions for immediate gains, or unintended changes in context. To ensure the evolution of the software, it is necessary to keep it managed. Identification and measurement are the first two stages of the management process; however, they are little explored in academic research in requirements engineering.

\noindent \textbf{Objective:} We aimed at investigating which evidence helps to strengthen the process of TD requirements management, including identification and measurement. 

\noindent \textbf{Method:} We conducted a Systematic Literature Review through manual and automatic searches considering 7499 studies from 2010 to 2020, and including 61 primary studies. 

\noindent \textbf{Results:} We identified some causes related to Technical Debt  requirements, existing strategies to help in the identification and measurement, and metrics to support the measurement stage. 

\noindent \textbf{Conclusion:} Studies on TD requirements are still preliminary, especially on management  tools. Yet, not enough attention is given to interpersonal issues, which are difficulties encountered when performing such activities, and therefore also require research. Finally, the provision of metrics to help measure TD is part of this work's contribution, providing insights into the application in the requirements context.  
\end{abstract}

%\includegraphics{figs/grabs.pdf}

%\item Research highlights item 1
%\item Research highlights item 2
%\item Research highlights item 3

\begin{keywords}
Technical Debt \sep Technical Debt Requirements \sep Identification  \sep Measurement \sep Systematic Literature Review
\end{keywords}

\maketitle

\section{Introduction}
In software development, even in well planned projects, some challenges can negatively influence their delivery and final quality, such as pressure from the customer to complete the software before the deadline, limited resources, or pressure from the market itself \citep{rios2018tertiary}. Challenged with this scenario, the development team needs to use alternative solutions to achieve the goals and tasks in the short term, often without considering the possibility of negatively impacting the software in the long term \citep{da2017using}.  In this context, quality problems can be observed in the project during or after its implementation. Thus, tasks that have been compromised need to be improved at some point throughout the project, and if not reached, can generate a phenomenon known as Technical Debt (TD)
\citep{cunningham1992wycash}.

TD refers to problems caused when software development tasks are pending or inefficiently executed
\citep{kruchten2012technical}. While these actions can provide benefits in the short term, such as increased productivity, there are risks to the project, hindering its evolution (\citeauthor{guo2016exploring}, \citeyear{guo2016exploring}; \citeauthor{rios2018most}, \citeyear{rios2018most}). With this, TD includes items usually controlled in a software project, such as unimplemented features. Also, it covers less visible aspects such as code smells and outdated documentation \citep{brown2010managing}.

Initially, TD had the focus on coding activities
\citep{cunningham1992wycash}, but in the advancement of investigations, the concept was expanded to cover the other phases of software development, for example, in requirements engineering
\citep{li2015systematic}. According to \citet{ernst2012role}, an inappropriate elicitation or analysis of requirements causes errors that increase the incidence of TD in software projects. In this context, the technical debt of requirements may occur intentionally, for example, in the case of consciously choosing not to execute the elicitation process strictly; or unintentionally, in cases where the requirements engineers are inexperience and may not have the skills needed to perform technical and long-term procedures
\citep{rios2019supporting}.

Regardless of how TD occurs, it is necessary to keep it managed to ensure the software's evolution and quality, avoiding a late discovery of its amplitude and consequently cost, causing the incidence of interest for correction \citep{alves}. Identification and measurement are the first two steps in the management process \citep{li2014architectural}. They are essential activities to know what type of TD exists, where it is located, and how to estimate its impact on the software \citep{alves2016identification}. However, they are considered the phases in which there is greater difficulty in achieving \citep{besker2018technical}. However, although professionals and researchers give much attention to the technical debt in recent years
(\citeauthor{rios2018tertiary}, \citeyear{rios2018tertiary}; \citeauthor{gama2019technical}, \citeyear{gama2019technical}), in the requirements engineering area, the process management, specifically in the identification and measurement of TD are still a gap to explore in academic research \citep{alves}.

In this context, the current work presents a Systematic Literature Review (SLR) to identify and give an overview of state of the art related to technical debt management in software requirements, specifically about the stages of identification and measurement of this type of TD. In the end, as main contributions, it is presented evidence and studies that shows the leading causes attributed to the emergence of the technical debt of requirements; strategies used to assist in its identification, as well as support metrics in the measurement stage; finally, identify gaps and opportunities for the development of new research, encouraging other researchers to continue research in the area. 

In addition to this section, the rest of this work is structured as follows: Section \ref{section2} presents the background; Section \ref{section3} contains the methodology used to perform the SLR; Sections \ref{section4} and \ref{section5} report and discuss the results obtained; Section \ref{section6} exposes the threats to the validity of this work; and finally, Section \ref{section7} contains the conclusions and future work. 

\section{Background} \label{section2}

This section goal is to present the main concepts that underpin the accomplishment of this work, as well as the explanation of the related works.

\subsection{Technical Debt}

According to \citet{seaman2011measuring}, TD is defined as immature or incomplete artifacts present in the software development life cycle, causing higher costs and low quality. The creation of these artifacts can accelerate development in the short term. However, low quality tends to generate expenses in the long run due to maintenance efforts used for corrections. And according to \citet{mcconnell2008managing}, technical debt can be categorized into two types:

\begin{itemize}
\item  \noindent \textbf{Unintentional TD}, which is ocurred in an involuntary and non-strategic way, many times caused when the activities are poorly planned, by the inexperience of the professionals or changes in the environment;

\end{itemize}

\begin{itemize}
\item \noindent \textbf{Intentional TD}, is motivated consciously and strategically, where professionals and teams make decisions to obtain short-term benefits, resulting from shortcuts, alternative solutions, tasks not performed or with the required level of detail.
\end{itemize}

Additionally, according to \citet{rios2018tertiary}, TD may be present in different activities and phases of the software development life cycle. With this, these same authors present a set containing 15 different types of technical debts identified. The Table
\protect\ref{Tab: Types_TD} presents each type and respective definition.

\begin{table} [pos=ht]
\caption{Types of Technical Debt.}\label{Tab: Types_TD}
\begin{tabular}{{p{0.75in}p{2.22in}}}
\hline
\multicolumn{1}{l}{\textbf{Type}} & \multicolumn{1}{l}{\textbf{Definition}} \\ \hline
Design              & \begin{tabular}[c]{@{}l@{}}Refers to TD discovered by analyzing the\\ source code and identifying violations of\\ principles of good object-oriented design.
\end{tabular}                                          \\ \hline
Code              & \begin{tabular}[c]{@{}l@{}}Refers to problems found in the source\\ code (violating best practices or coding\\ rules) that negatively affect its readability\\ and make it difficult to maintain. \end{tabular}                                                              \\ \hline
Architecture         & \begin{tabular}[c]{@{}l@{}} Refers to problems found in the product\\ architecture, which affect the architecture\\ requirements. Generally, this TD is the\\ result of initial solutions below the ideal,\\ compromising internal aspects of quality.
\end{tabular}                                                             \\ \hline
Test              & \begin{tabular}[c]{@{}l@{}}Refers to problems found in testing\\ activities that affect their quality. 
\end{tabular}                                                               \\ \hline
Documentation       & \begin{tabular}[c]{@{}l@{}}Refers to the problems found in software\\ project documentation. \end{tabular}                                           \\ \hline
Defect            & \begin{tabular}[c]{@{}l@{}}Refers to known defects, usually identified\\ by test activities or the user. The\\ development team agrees to correct them,\\ but due to competing priorities and limited\\ resources, they will be delayed. \end{tabular}                                                                      \\ \hline
Infrastructure      & \begin{tabular}[c]{@{}l@{}}Refers to infrastructure problems that, if\\ present in the software organization, delay\\ or hinder development activities. Such TD\\ negatively affects the team's ability to\\ produce a quality product. \end{tabular}                                               \\ \hline
Requirements           & \begin{tabular}[c]{@{}l@{}}Refers the distance between the optimal\\
requirements specification and the actual\\ system implementation.\end{tabular}                                                                 \\ \hline
People             & \begin{tabular}[c]{@{}l@{}}Refers to people issues that, if present in\\ the software organization, can delay or\\ hinder some development activities.\end{tabular}                                                 \\ \hline
Build          & \begin{tabular}[c]{@{}l@{}}Refers to issues that make the build task\\ harder, and unnecessarily time consuming.\end{tabular}                                                         \\ \hline
Process            & \begin{tabular}[c]{@{}l@{}}
Refers to inefficient processes, e.g. (the\\ projected process may not be appropriate).
\end{tabular}                                                              \\ \hline
Automation           & \begin{tabular}[c]{@{}l@{}}Refers to the work involved in the\\ automation of functionality tests developed\\ to support continuous integration and\\ faster development cycles.
 \end{tabular} \\ \hline
Usability         & \begin{tabular}[c]{@{}l@{}}Refers to inappropriate usability decisions\\ that will need to be
adjusted later.\end{tabular}                                                                              \\ \hline
Service             & \begin{tabular}[c]{@{}l@{}}Refers to inappropriate web services that\\ lead to incompatibility between service\\ features and application requirements.  \end{tabular}                                       \\ \hline
Versioning      & \begin{tabular}[c]{@{}l@{}}Refers to problems in source code \\versioning, such as unnecessary code forks.
\end{tabular}                                                                                \\ \hline
\end{tabular}

\end{table}

\subsection{Technical Debt Requirements}

Requirements engineering is one of the areas of software engineering that has as its objectives to include the use and analysis of techniques and activities to obtain, specify and document a set of requirements that meet the needs of stakeholders with a high-quality product
\citep{vazquez2016engenharia}.  The requirements describe the software, its behaviour, its functionality, constraints and all other attributes. However, it is a tricky step, which often does not receive proper attention from stakeholders (customers, analysts, users, developers)
\citep{wiegers2013software}.

According to \citet{van2008software}, stages and tasks of requirements engineering when performed inadequately, can cause problems that affect the development of the software, such as low quality elicitation, incomplete or outdated requirements together with other existing problems, are real examples of technical debt. For \citet{ernst2012role}  and \citet{brown2010managing}, the TD of requirements is related to the distance between the ideal value of the specification of requirements and the actual implementation of the system, which is a consequence of strategic decisions for immediate gains, or unintended changes in context, which lead to future costs.

This type of technical debt belongs to the exchanges made regarding which requirements the development team needs to implement or how to implement them, and according to \citet{abad2015using}, can be defined as trade-offs during the specification of requirements. In this context, the requirements that are partially implemented, not specified in a satisfactory manner, poorly prioritized or developed without considering their dependencies and relationships, represent errors that increase the incidence of TD, leading to increased interest and effort needed for the correction if this type of technical debt is not identified and managed at the correct time (\citeauthor{ernst2012role}, \citeyear{ernst2012role}; \citeauthor{abad2015using}, \citeyear{abad2015using}).

Recently, the work of \citet{lenarduzzi2019towards} defined the TD of requirements in three types:

\begin{itemize}
\item \textbf{Type 0: Incomplete Users’ needs}
\end{itemize}
Represents TD incurred by neglecting the needs of stakeholders or a specific group of stakeholders.  For example, in the case of consumer-centric systems such as mobile applications, TD is incurred when users' needs expressed in feedback channels are forgotten. The authors also present that this type of TD of requirements can be quantified as the proportion between the user needs that have already been elicited and all possible needs, including neglected ones. The principal one is the cost of obtaining all the user's remaining needs, and the interest is the cost associated with the risk of missing an important need. In other words, a requirements engineer needs to decide whether it is worth spending time identifying additional user needs, taking into account, for example, the current development stage of the software associated with implementing a requirement detected later.

\begin{itemize}
\item \textbf{Type 1: Requirement smells}
\end{itemize}
Represents the TD incurred when linguistic constructions may indicate a violation of ISO/IEC/IEEE 29148:2018, which references the quality of requirements. These smells also exist for other requirements documentation approaches, e.g. UML. If these smells are not removed, the requirement may be implemented incorrectly, making it difficult to reuse and evaluate. In this sense, the authors state that requirement smells also need to be reimbursed similar to code smells. With this, the principal can be quantified as the cost to correct them and the interest as the negative impact  on the stages of software development with which they are associated.

\begin{itemize}
\item \textbf{Type 2: Mismatch implementation}
\end{itemize}
Represents the TD incurred when developers implement a solution to a requirements problem. Thus, an incompatibility is identified between the stakeholders' objective framed during the specification of requirements and the actual implementation of the system. According to the authors, a way to identify this type of requirements technical debt can be based on approaches to traceability between the requirements specification and source code, such as RE-KOMBINE \citep{ernst2012role} (will be presented in Subsection \ref{rq2}). Finally, this third type of requirements technical debt is quantified as the cost of comparing the current software implementation with the set of possible changes (Principal), additionaly the performance of the selected change (Interest).

\subsection{Technical Debt Management}

According to \citet{li2015systematic}, the management of TD is an important step to achieve good quality in the development and maintenance of the software, since most of the debts are often not managed. By  \citet{tom2013exploration}, it is necessary to define processes that can track these technical debts, so that later, decisions can be based on the problem identified.  Also, recent research shows that knowing the existence of technical debt influences the behaviour of the team, i.e. applying the best techniques of identification and measurement, for example, can
significantly improve software development practices \citep{tonin2018technical}. 

The management process includes activities used to control and reduce the technical debt in a software project. In this context, with the inclusion of different techniques, tools and evidence, companies aim to reduce and prevent shortcuts and solutions that do not achieve the expected success \citep{li2015systematic}. However, most of the TD items are inadequately managed, thereby further increasing the risk of high maintenance costs \citep{tonin2017effects}. Therefore, it is appropriate to find the best ways to ensure that the TD achieve proper management, facilitating decision-making on future activities \citep{alves}.

In the work of \citet{li2014architectural}, a technical debt management method was proposed, containing five steps: identification, measurement, prioritization, repayment and monitoring, which are described below.
 \newline

\noindent \textbf{1. Identification:} the process of visualizing the technical debt, identifying its causes and other attributes present in software development that led to its existence. This activity is crucial for the proper management of TD.

\noindent \textbf{2. Measurement:} analyzes and quantifies the costs and efforts required to assist in decision making regarding technical debt reimbursement. 

\noindent \textbf{3. Prioritization:} organize the payment of technical debts in relation to importance, analyzing factors such as technical issues and financial impacts.

\noindent \textbf{4. Repayment:} regarding the partial or total payment of the technical debt, avoiding postponing it if it could negatively affect the project.

\noindent \textbf{5. Monitoring:} validates whether the technical debt is being diluted, delayed, or continues to cause costs. 

\subsection{Related Works}

This subsection presents the works related to the objectives proposed by this study. They are listed in chronological order and can be identified in Table \ref{Tab: work}.  Next, the details of each study are presented, and a comparative analysis of the differences concerning this study.

\begin{table} [h]

\caption{Related Works}\label{Tab: work}

\begin{tabular}{cl}
\hline
\textbf{Work}                                                      & \multicolumn{1}{c}{\textbf{Goal}}                                                                                       \\ \hline
\begin{tabular}[c]{@{}c@{}}Alves et al.\\ (2016)\end{tabular}      & \begin{tabular}[c]{@{}l@{}} TD management strategies and\\ TD taxonomy\end{tabular}                                       \\ \hline
\begin{tabular}[c]{@{}c@{}}Nascimento\\ et al. (2018)\end{tabular} & \begin{tabular}[c]{@{}l@{}}Investigate and conceptualize\\ requirements smells\end{tabular}                              \\ \hline
BenIdris (2020)                                                    & Analyze TD in empirical studies                                                                                          \\ \hline
\begin{tabular}[c]{@{}c@{}}Wang and Huang\\ (2020)\end{tabular}    & \begin{tabular}[c]{@{}l@{}}Conceptualize requirements TD\\ and find approaches to manage it\end{tabular}                \\ \hline
\begin{tabular}[c]{@{}c@{}}Lenarduzzi et al.\\ (2021)\end{tabular} & \begin{tabular}[c]{@{}l@{}}Identify TD prioritization tools,\\ strategies, processes and factors\end{tabular}            \\ \hline
Our work                                                      & %\begin{tabular}[c]{@{}l@{}}Identify strategies and metrics to \\ identify and measure TD of \\ requirements\end{tabular} \\ \hline
\end{tabular}

\end{table}

The work of \citet{alves2016identification} was aimed at conducting an Systematic
Mapping Study (SMS) to analyze which strategies proposed to help manage TD in software projects and analyze their main types. The search process executed automatically, recovering searches in the period 2010 to 2014, and at the end, 100 studies were considered. Among the results, they proposed an initial taxonomy of the TD types and a list of existing strategies for identification and management. Finally, a current state-of-the-art analysis identified gaps where new research efforts could be invested.

In work proposed by \citet{nascimento2018requirements}, a SMS was conducted to investigate evidence on the subject of requirements smells, thus helping in their understanding and assisting researchers in future studies. The search process was automatically executed, recovering research with a publication date from January 2013 to March 2018, and at the end, 41 studies were considered. Among the results, it was identified that the concept had gained visibility in recent years and the development and existence of support by tools.

In work proposed by \citet{benidris2020investigate}, a SMS was executed to identify and analyze TD in empirical studies published from 2014 to 2017. The search process was carried out in an automated way, and in the end, 100 studies were considered. Among the main results, the presentation of the most common indicators and evaluators to identify and evaluate TD and, the identification of tools and strategies that help to investigate and estimate.

The work of \citet{wang2020identification} had as a proposal to investigate the current state on the TD of requirements, to be able to present a precise definition on this type of TD. To achieve this goal, they conducted an SMS, and a survey. Among the results, ten measurement techniques were identified and suggestions from software professionals about the detection of this TD.  Finally, they discovered that academia and industry are different in managing this TD and encouraging both sides to work collaboratively.

Finally, the paper of \citet{lenarduzzi2021systematic} investigated which approaches to prioritizing TD have been proposed in software engineering research and industry. To do this, they conducted an SLR, which in the end included 44 primary studies. Among the results, they observed that research on TD prioritization is preliminary and that there is no consensus on which factors are essential and how to measure them.  And in the end, they will propose a mind map that can help software professionals during TD prioritization.

The works cited and the present article is related because they seek ways to understand and manage TD. But in contrast with the works mentioned, this study conducted two types of search (manual and automatic). Also, we use snowballing method as a compliment. This generates more research sources to be considered. In the case of the work of  \citet{nascimento2018requirements} and \citet{wang2020identification}, the proposal resembles by investigating requirements smells and TD requirements. Although, they did not address specific contents on the causes for their emergence and metrics for measurement.  The works of  \citet{alves2016identification} and \citet{benidris2020investigate} focused on TD in general,   gathering evidence to help in its management. Finally, the paper of \citet{lenarduzzi2021systematic} analyzed a specific step (prioritization) of the TD management process but differed by not addressing evidence focused on its identification and measurement.

\section{Systematic Literature Review} \label{section3}

\begin{figure}[ht]
	\centering
		\includegraphics[scale=.552]{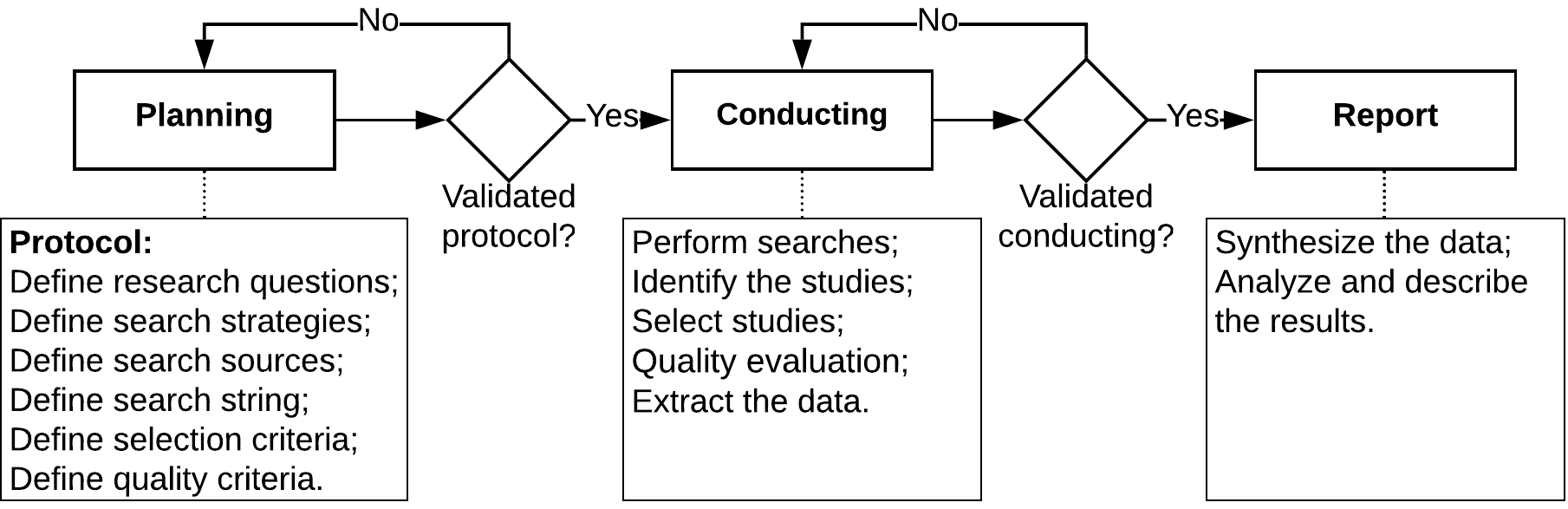}
	\caption{Process of Conducting Systematic Literature Review.}
	\label{FIG:1}
\end{figure}

The SLR conducted in this work was based on the method proposed by  \citet{kitchenham2007guidelines}. According to the authors, a systematic approach is pre-defined using a protocol and procedures to identify, evaluate and interpret the relevant evidence available in Primary Studies (P), related to one or more research questions. The process of this SLR is presented in Figure \ref{FIG:1}.

\subsection{Planning}
\subsubsection{Research Questions}

This work's main objective is to provide evidence to assist in the identification and measurement of TD requirements in software development.  To understand this objective, the following Research Question (RQ) was defined: "How to assist in the identification and measurement of the technical requirements debt in software development?". To answer this question, we have derived it into for sub-questions:
 \newline

\noindent \textit{\textbf{RQ1:} What has caused the technical debt of requirements in software development?}

Identifying a TD is not only about understanding how and where it occurred—but also analyzing the causes that led to its occurrence. The answers to this question will help you understand the causes of the emergence of TD requirements, aiding in their identification and prevention.
 \newline

\noindent \textit{\textbf{RQ2:} What strategies are proposed to help identify and measure the technical debt requirements in software projects?}

As important as managing TD items in a project is implementing efficient and effective management strategies for such activities. Answering this question may help practitioners in the selection of strategies and tools already available. \newline

\noindent \textit{\textbf{RQ3:} What metrics are being used to assist in the process of measuring the technical requirements debt?}

This question aims to identify a set of metrics that are considered valid and provide more accurate estimates during the measurement of a TD. In addition, it seeks to understand the main variables considered in this step, such as the principal and interest of the technical debt.
 \newline

\noindent \textit{\textbf{RQ4:} What difficulties are pointed out during the management of technical debt requirements in software development?}

The answer to this question will provide a list of difficulties encountered by software professionals when working with requirements TD in practice. Subsequently, this evidence will be used to develop future research, directing new efforts that will support the management of TD.

%\todo[inline]{Valentina: add motivation for each RQ}

\subsubsection{Sources and Search String}

The research for primary studies was initially done through manual and automatic searches in specialized and renowned scientific-academic sources and digital libraries in Computer Science and the subjects related to the objective of this work. The Table \ref{Tab: Sources} presents the search sources used.

\begin{table} [pos=h]

\caption{Sources and Digital Libraries Used.}\label{Tab: Sources}
\begin{tabular}{{p{3.0in}}}
\hline
\textbf{Manual Search}                                                                                                    \\ \hline
Information and Software Technology (IST)                                     \\ \hline
International Journal of Software Engineering and Knowledge Engineering (SEKE)              \\ \hline
International Requirements Engineering Conference (RE)                                                     \\ \hline
International Workshop on Managing Technical Debt (MTD) \\  International Conference on Technical Debt (TechDebt)                                                           \\ \hline
Journal of Systems and Software (JSS)                                            \\ \hline
\textbf{Automatic Search}                                                                                                 \\ \hline
ACM Digital Library                                                                                                    \\ \hline
IEEE Digital Library (IEEEXplore)                                                    \\ \hline
Science Direct                                                                                                    \\ \hline
SCOPUS                                                                                                       \\ \hline
SpringerLink                                                                                                  \\ \hline
\end{tabular}
\end{table}

Searching for relevant results among digital libraries in the automatic search, a search string was formed based on two criteria: (i) higher number of results recovered from digital libraries and (ii) studies strongly related to the search theme. We would like to highlight that our search string is based on the definitions of the secondary studies of: \citet{behutiye2017analyzing} related to TD; the study of \citet{saha2012systematic} on software requirements; finally, the study of \citet{riaz2009systematic} adapting the terms related to measurement and metrics. With this, the following search string was defined:

\begin{center}

\textit{(``technical debt'' OR ``technical debit'' OR ``design debt'' OR ``debt metaphor'') \newline  AND \newline (``requirement'' OR ``requirements'' OR ``requirement engineering'' OR ``software requirements'' OR ``user story'' OR ``measurement'' OR ``metrics'' OR ``measure'' OR ``measurement metrics'')}

\end{center}

\subsubsection{Selection Criteria}

Inclusion (I) and Exclusion (E) criteria were defined to assist the primary studies' selection process. The criteria can be observed in the Tables \ref{Tab: Inclusion} and \ref{Tab: Exclusion}.

\begin{table} [pos=h]
\caption{Inclusion Criteria.}\label{Tab: Inclusion}
\begin{tabular}{l}
\hline
\textbf{I1: }Studies published between 2010 and 2020                                                                                                           \\ \hline
\textbf{I2: }Studies wrote in English                                                                                                    \\ \hline
\begin{tabular}[c]{@{}l@{}} \textbf{I3: }Studies related to the identification and measurement\\ of technical debt requirements in software development\end{tabular} \\ \hline
\end{tabular}
\end{table}

\begin{table} [pos=h]
\caption{Exclusion Criteria.}\label{Tab: Exclusion}
\begin{tabular}{l}
\hline

\textbf{E1:} Studies that are not related to the research questions              \\ \hline
\textbf{E2:} The study is not accessible                                                 \\ \hline
\textbf{E3:} Secondary and tertiary studies                                              \\ \hline
\textbf{E4:} The study is a copy or an older version of another \\study already considered \\ \hline
\textbf{E5:} Studies published before 2010 \\ \hline
\end{tabular}
\end{table}

\subsubsection{Quality Assessment Criteria}

The quality evaluation of the studies was performed to ensure that the final selection list included the most relevant to this work's objective. For this purpose, the Quality Criteria (Q) proposed by \citet{dyba2007applying} were used, which are evaluated in the following quality guidelines: \newline

\noindent \textit{\textbf{Reporting:}} the quality of the logic of the objectives and the context of the study;

\noindent \textit{\textbf{Credibility:}} the rigor of the research methods used to establish the validity of the data collection tools and analysis;

\noindent \textit{\textbf{Rigor:}} evaluates the credibility of study methods to ensure that they are valid and meaningful;

\noindent \textit{\textbf{Relevance:}} address the relevance of the study to the software industry and the research community. \newline 

In this process, all studies were read entirely, and at the end, a score was assigned to each criterion presented in the Table \ref{Tab: Quality}. The possible scores were: \newline 

\noindent[0] The study does not meet the quality criteria; 

\noindent[1] The study fully meets the quality criteria.

\begin{table} [pos=ht]
\caption{Quality Criteria.}\label{Tab: Quality}
\begin{tabular}{{p{2.51in}p{0.5in}}}
\hline

\begin{tabular}[c]{@{}l@{}}Q1: Is the research related to the identification\\ and measurement of TD requirements? \end{tabular} & Reporting                             \\ \hline
Q2: Are the objectives clearly defined?                                                                                                                            & Reporting                            \\ \hline
\begin{tabular}[c]{@{}l@{}}Q3: Is there an adequate description of the\\ context in which the research was carried out?\end{tabular}                                               & Reporting                              \\ \hline
Q4: Is the application domain clearly expressed?                                                                                                                       & Reporting                           \\ \hline
\begin{tabular}[c]{@{}l@{}}Q5: Was the research design appropriate to\\ meet the research objectives?\end{tabular}                                             & Rigor                                  \\ \hline
Q6: Was the data analysis sufficiently rigorous?                                                                                                                 & Rigor                                  \\ \hline
Q7: Is the type of research conducted clearly expressed?                                                                                                               & Rigor                                  \\ \hline
Q8: Are the results clearly described?                                                                                                                             & Credibility                          \\ \hline
Q9: Is it possible to identify the place of publication of the research?                                                                                                          & Credibility                         \\ \hline
Q10: Is the contribution clearly expressed?                                                                                                                           & Credibility                        \\ \hline
Q11: Does the research make it clear who \\contributes?                                                                                                                          & Relevance                            \\ \hline

\end{tabular}
\end{table}

It was also defined following \citet{dyba2007applying}, that if the primary study did not meet Q1, it would be excluded. Similarly, if Q2, together with Q3, were not completed, the study would be removed.  Also, following the recommendations of \citet{lima2019metodologias}, a minimum score of 6 points was required for the study to be considered in the final list of this SLR, i.e. it had to achieve more than 50\% of the criteria.

\subsection{Conducting}

In the second stage of this SLR, the manual search was initially performed by the primary studies through access to the annals of the search sources. In the automatic search, the studies were identified by applying the search string to the digital libraries. The primary search resulted in 6508 primary studies. The inclusion and exclusion criteria were used, resulting in 250 selected studies. These had their titles, abstracts, and keywords analyzed. During the quality assessment process, the resulting studies were fully read to identify which ones met the quality criteria and satisfactorily answered the research questions, resulting in 61 studies. 

\begin{figure*}[ht]
	\centering
		\includegraphics[scale=.24]{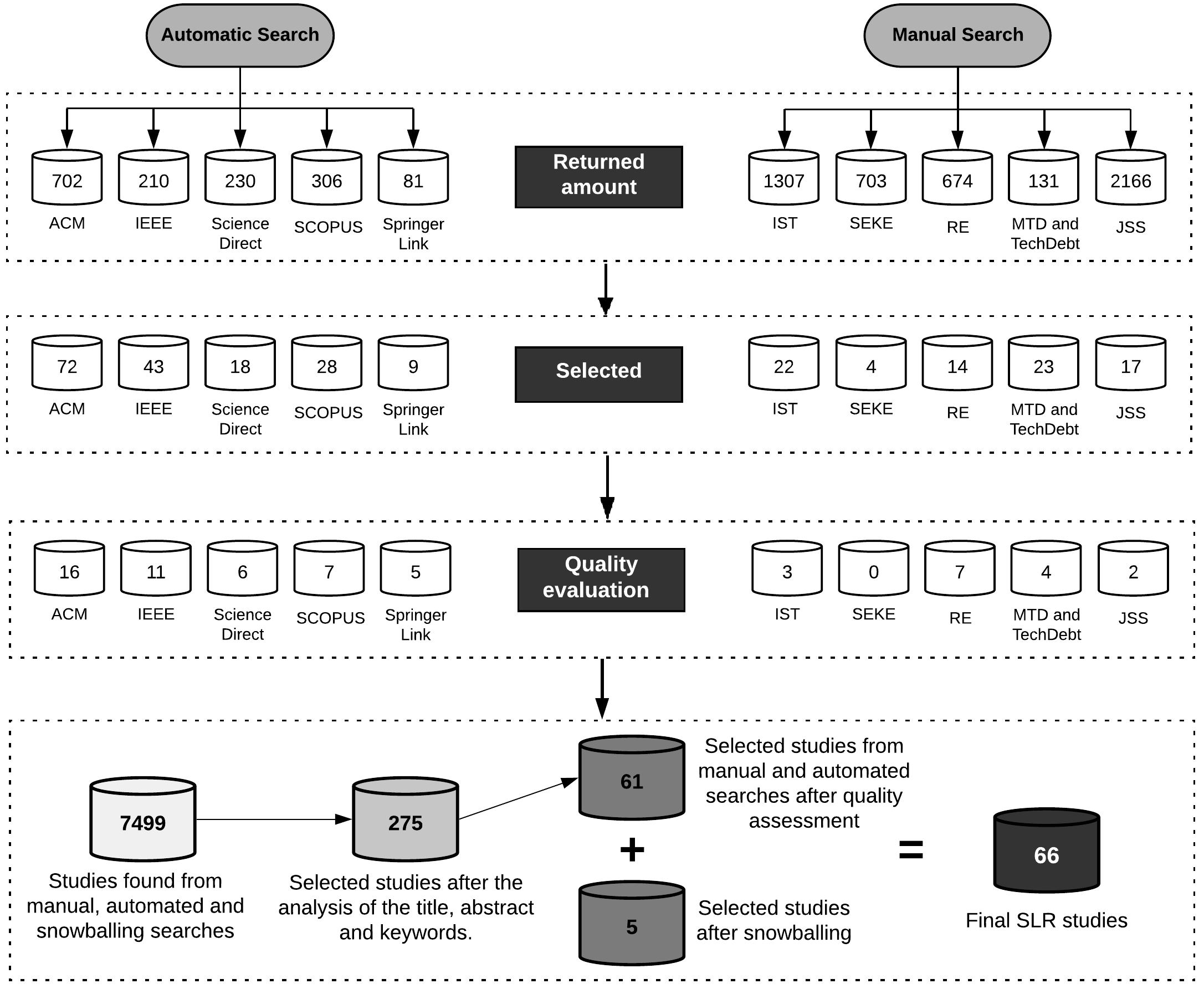}
	\caption{Process of Conducting Systematic Literature Review.}
	\label{FIG:2}
\end{figure*}

In the sequence, to complement the evidence already identified and to guarantee the integral inclusion of studies related to this work's objective, the conduction of the snowballing method was included \citep{wohlin2014guidelines}. At this point, a total of 991 references cited in the 61 primary studies included were analyzed. In the end, 25 studies were selected through the references and, after full reading and quality assessment, five studies were including, resulting in the final version of this SLR (66 studies), as detailed in Figure \ref{FIG:2}.

After data extraction phase, the data were extracted, aiming to obtain the information needed to answer the research questions, and a spreadsheet was used for this process. The Table \ref{Tab: Extraction} presents the data extracted from the 66 studies. 

% Next, the spreadsheet that presents the conducting of this SLR is available and detailed online at this.

\begin{table} [pos=ht]
\caption{Information Extracted from Primary Studies.}\label{Tab: Extraction}
\begin{tabular}{ll}
\hline
\textbf{Information}                                  & \textbf{Research Question} \\ \hline
Title                                                 & Overview                   \\ \hline
Author                                                & Overview                   \\ \hline
Year of publication                                   & Overview                   \\ \hline
Place of publication                                  & Overview                   \\ \hline
Research method                                       & Overview                   \\ \hline
Cause attributed to the emergence\\ of requirements TD  & RQ1                        \\ \hline
Proposed strategy to identify the\\ requirements TD     & RQ2                        \\ \hline
Proposed strategy to measure the\\ TD of requirements   & RQ2                        \\ \hline
Metric used to measure the TD of\\ requirements         & RQ3                        \\ \hline
Difficulty reported when managing\\ the requirements TD & RQ4                        \\ \hline
\end{tabular}
\end{table}

The process of interpretation of the results was initiated from the extracted data, elaborating tables, graphs, and networks to present the identified information to answer the research questions. We would like to emphasize that this procedure was performed with the qualitative analysis tool Atlas.ti\footnote{https://atlasti.com/}. The final list of analyzed studies is available in the Appendix A. Finally, to avoid research bias, the entire process and analysis of SLR was executed, discussed and reviewed by all the authors of this work.

\subsection{Verifiability and Replicability}
%\todo[inline]{Please add the link to the rawdata}
In order to allow replication and extension of our work by
other researchers, we prepared a replication package~\footnote{\label{package} http://bit.ly/StudiesPrimary} for this
study with the complete results obtained.

\section{Results} \label{section4}

This section reports the evidence and information found in the systematic literature review. Sub-questions between the following subsections present the quantitative and qualitative results and their analysis. But initially, an overview of the 66 primary studies analyzed in this systematic review is provided.

\subsection{Overview of Primary Studies}

\begin{figure}[ht]
	\centering
		\includegraphics[scale=.58]{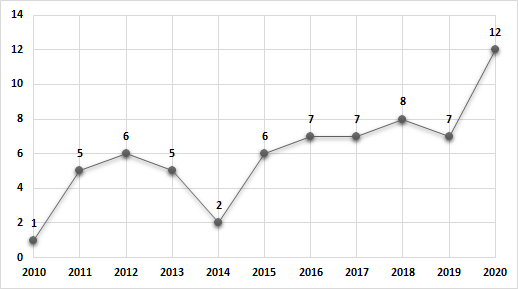}
	\caption{Publications Evolution in the Last 10 Years.}
	\label{FIG:3}
\end{figure}

A total of 66 studies were published considering the period from 2010 to 2020. As shown in Figure \ref{FIG:3}, only 19 studies were published by 2014, while almost 71\% of the studies were published from 2015. With this, it is possible to identify that the number of publications and research in the area has increased in recent years. Highlight to the year 2020 with 12 publications, in contrast to the year 2014 with two publications, confirming results of other secondary studies published recently, as the year with fewer publications on the subject of TD  (\citeauthor{becker2018trade}, \citeyear{becker2018trade};  \citeauthor{lenarduzzi2021systematic}, \citeyear{lenarduzzi2021systematic}). 

When considering the distribution of studies based on location and type of search, a relatively high percentage was identified
of studies (75\% or 49 studies) published and attached on digital bases. While only 25\% of studies (17 studies) have been published in journals and manual databases. This analysis can be better visualized previously in Figure \ref{FIG:2}.

Then, as one of the objectives of this work it was to analyze studies related to the TD of requirements, as well as studies that examined the process of measurement and the supply of metrics, so that it could verify the possibility of using, if necessary adapt, these proposed metrics in the context of requirements. It was identified that 48 studies (approximately 72\%), were related and provided evidence focused on TD of requirements. At the same time, 18 studies (approximately 28\%) were associated with analyzing the process of measuring a TD and offering metrics.

Soon after, it was identified that the 66 primary studies selected were written by 130 different authors, showing a broad interest in this subject. However, it was found that only 13 researchers were involved in at least three articles each. In the following sequence, the respective authors are presented in the Table \ref{Tab: author} in order of representativeness.

\begin{table} [pos=h]
\caption{Authors with Greater Representation among Primary Studies.}\label{Tab: author}
\begin{tabular}{cl}
\hline
\textbf{Number of Primary Studies} & \multicolumn{1}{c}{\textbf{Author's name}}                                                                    \\ \hline 10                                  & \begin{tabular}[c]{@{}l@{}}Carolyn Seaman\\ Rodrigo O. Spínola\end{tabular}   \\ \hline
8                                  & Manoel Mendonça                                                \\ \hline
7                                  & Antonio Martini     \\ \hline
6                                 & Jan Bosch                                                                                              \\ \hline
5                                  & \begin{tabular}[c]{@{}l@{}}Yuepu Guo\\ Terese Besker \end{tabular}                                \\ \hline
4                                  & \begin{tabular}[c]{@{}l@{}}Forrest Shull\\ Alexander Chatzigeorgiou\\Nicolli Rios

\end{tabular}               \\
\hline
3                                  & \begin{tabular}[c]{@{}l@{}}Valentina Lenarduzzi\\ Paris Avgeriou
\\Nico Zazworka

\end{tabular}               \\
\hline
\end{tabular}
\end{table}

\begin{figure}[ht]
	\centering
		\includegraphics[scale=.78]{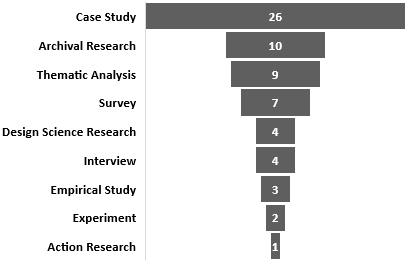}
	\caption{Applied Research Method in Primary Studies.}
	\label{FIG:4}
\end{figure}

Finally, the analysis of primary studies was performed about the research method applied, which was based on the classification presented in the work of \citet{molleri2019cerse}. Thus, as shown in the Figure \ref{FIG:4}, the \textit{case study} stood out in a total of 26 publications, following \textit{archival research}
 (10), which investigates the data through documental analysis and reports, for example. In the sequence, \textit{thematic analysis}
 (9), \textit{survey} (7), 
 \textit{design science research} and
 \textit{interview}  (each with four publications), \textit{empirical study } (3), \textit{experiment} (2), and finally \textit{action research} (1).
 
\subsection{RQ1: What has caused the technical debt of requirements in software development?}

Second \citet{li2015architectural}, one of the variables that helps in the identification of a TD, are the causes that caused its emergence. However, \citet{rios2018tertiary} reports that this is a topic (causes for TD insertion in projects) that remains little explored in academic research. In this sense, this question objective was to identify the main causes of the emergence of the TD of requirements, to facilitate its identification, presenting the main indicators for its occurrence.
 
After the analysis, 33 causes (codes) were identified and, in the sequence, the level of grounded (quantity of citations in studies) and density (amount of association with other codes) for each one was verified. For space reasons, the 15 causes of greater representativity considering these two variables are presented in the Table \ref{Tab: RQ1}. The other causes are reported throughout this subsection and, all of which are detailed and defined online at the following link\footnote{http://bit.ly/DetailingRQ1}.

\begin{table*} [pos=h]

\caption{Leading Causes Attributed to the Emergence of Technical Requirements Debt.}\label{Tab: RQ1}
\begin{tabular}{l c c c l}
\hline

\textbf{Cause}                                    & \textbf{Grounded} & \textbf{Density} & \multicolumn{1}{l}{\textbf{Amount}  } & \multicolumn{1}{c}{\textbf{Primary Studies}}     \\ \hline
Low level of detail in requirements documentation & 10                 & 3                & 13     & \begin{tabular}[c]{@{}l@{}}P74, P91, P154, P183, P192, P209, \\P214, P229, P233, P275 \end{tabular}                              \\ \hline

Ambiguous requirements                            & 7                 & 3                & 10                & \begin{tabular}[c]{@{}l@{}}P101, P125, P134, P135, P247,\\ P249, P263 \end{tabular}                    \\ \hline
Non-definition of non-functional requirements     & 9                 & 1                & 10            & \begin{tabular}[c]{@{}l@{}}P74, P81, P91, P196, P209, P214,\\P216, P233, P273 \end{tabular}                             \\ \hline
Vague and incomplete requirements                 & 7                 & 2                & 9                           & \begin{tabular}[c]{@{}l@{}}P125, P133, P154, P165, P196,\\ P209, P263 \end{tabular}  

\\ \hline

Lack of communication with the customer           & 6                 & 2                & 8                           & \begin{tabular}[c]{@{}l@{}}P9, P91, P117, P154, P214, P229 \end{tabular}                 \\ \hline

Shortcuts and alternative solutions                         & 4                 & 3                & 7   & \begin{tabular}[c]{@{}l@{}}P11, P13, P111, P117 \end{tabular}                                                \\ \hline

Schedule pressure                                 & 5                 & 1                & 6                          & \begin{tabular}[c]{@{}l@{}}P81, P101, P154, P161, P183 \end{tabular}               \\ \hline

Clients do not reflect what they want             & 4                 & 2                & 6                     & \begin{tabular}[c]{@{}l@{}}P101, P132, P167, P247 \end{tabular}                    \\ \hline
Lack of experience               & 5                 & 1                & 6         & \begin{tabular}[c]{@{}l@{}}P8, P167, P183, P229, P249 \end{tabular}                                \\ \hline
Inadequate prioritization of requirements         & 4                 & 1                & 5              & \begin{tabular}[c]{@{}l@{}}P25, P74, P91, P132 \end{tabular}                           \\ \hline
Inadequate writing and grammar                    & 3                & 1                & 4               & \begin{tabular}[c]{@{}l@{}}P133, P225, P263 \end{tabular}                          \\ \hline
Poorly planned interviews                         & 1                 & 3                & 4                  & \begin{tabular}[c]{@{}l@{}}P167 \end{tabular}                       \\ \hline
Lack of a script                         & 1                 & 2                & 3                  & \begin{tabular}[c]{@{}l@{}}P167 \end{tabular}                       \\ \hline
Inaccurate or complex requirements                        & 1                 & 1                & 2                  & \begin{tabular}[c]{@{}l@{}}P274 \end{tabular}                       \\ \hline
Inadequate elicitation                        & 1                 & 1                & 2                  & \begin{tabular}[c]{@{}l@{}}P25 \end{tabular}                       \\ \hline
\end{tabular}
\end{table*}

In the sequence, as explained above, the analysis of the results was supported by the qualitative tool Atlas.ti. This tool was idealized by \citet{muhr1991atlas}, based on Grounded Theory for its development. Among its many features is the possibility to build states of the art, multimedia analysis of videos, statistical treatment of data, analysis of surveys and, database coding. because of that, many researchers from different areas have used Atlas.ti in their research.

Among the analysis options that the tool offers, there is the possibility of creating networks, which would graphically represent the relationship between the codes identified, associating those that can cause or influence the existence of others. Thus, four networks were created for this issue, and the 33 causes are illustrated below.

As illustrated in the Figure \ref{FIG:5}, the first network presents 14 causes associated with the emergence of the TD of requirements. It becomes possible to notice, for example, that not properly planned interviews are part of an inadequate elicitation, as well as are associated with the lack of a script. This often causes clients not to reflect what they want, causing ambiguous requirements. 
Furthermore, it is observed that the absence of details in the documentation may lead to vague and incomplete requirements, resulting from their low definition and prioritization, as in the case of prioritizing requirements that do not offer greater value to the client.

 \begin{figure}[ht]
	\centering
		\includegraphics[scale=.762]{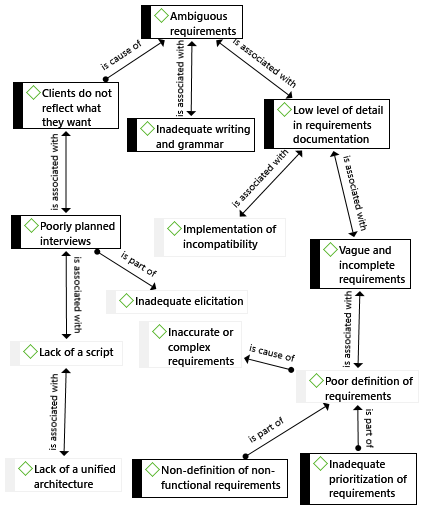}
	\caption{Association of the Causes of the Emergence of the TD Requirements (Network 1).}
	\label{FIG:5}
\end{figure}

\begin{figure}[ht]
	\centering
		\includegraphics[scale=.71]{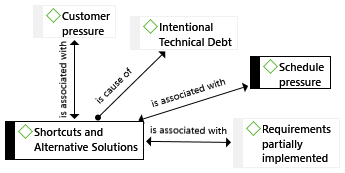}
	\caption{Association of the Causes of the Emergence of the Intentional Requirements TD (Network 2).}
	\label{FIG:6}
\end{figure}

The second network, presented in Figure \ref{FIG:6}, is related to the causes attributed to the emergence and identification of intentional requirements' technical debt. With this, it is possible to see that it is caused when professionals and software teams consciously choose to take shortcuts and alternative solutions involving activities related to software requirements. This cause can be directly influenced by schedule pressure and pressure from the client itself, which is considered by \citet{spinola2013investigating} the root cause of most technical debts.

\begin{figure}[ht]
	\centering
		\includegraphics[scale=.73]{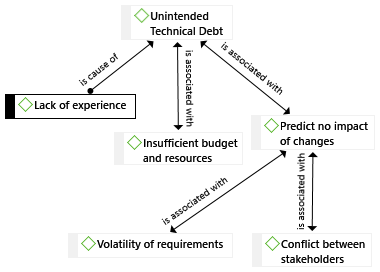}
	\caption{Association of the Causes of the Emergence of the Unintentional Requirements TD (Network 3).}
	\label{FIG:7}
\end{figure}

The causes associated with the emergence and identification of unintentional technical debt requirements were associated and presented in Figure \ref{FIG:7}. With this, it is possible to see that this TD can be caused by the lack of experience of the professionals and the insufficient amount of budget and human resources available in software projects.

Also, unintentional technical requirements debt may be caused by the difficulty predicting change impacts, i.e., predicting possible future updates or changes in requirements. This cause is associated with existing conflicts between stakeholders, as well as, the volatility of the requirements, i.e., the changes and updates that will occur throughout the project, since in the initial phase the requirements are not yet well defined and most of the time some details of the specification are only known during the implementation of the system.

\begin{figure}[ht]
	\centering
		\includegraphics[scale=.79]{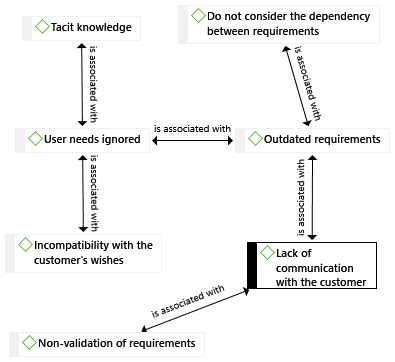}
	\caption{Association of the Causes of the Emergence of the TD Requirements (Network 4).}
	\label{FIG:8}
\end{figure}

Finally, the latest causes associated with the emergence of the technical requirements debt are presented in Figure \ref{FIG:8}. For example, it is possible to see that the non-validation of requirements is related to the absence of a review of the client's requirements. Sometimes, not enough attention is given to a detailed review of the requirements specification about their quality and domain-specific content.

In addition to this cause, the non-validation of requirements is mostly caused by a lack of communication with the customer. Thus, this cause can generate outdated requirements, i.e. they refer to cases where they were developed at an appropriate level of quality (in the first versions of the system). Subsequently, the specifications are not updated with new requirements or changes to those already existing.

\subsection{RQ2: What strategies are proposed to help identify and measure the technical debt requirements in software projects?} \label{rq2}

This question objective was to identify and present the strategies that already exist to assist in the identification and measurement of the technical requirements debt. In total, 18 strategies were identified and shown in the Table \ref{Tab: RQ2} considering representativeness among the primary studies. In the sequence, each strategy is detailed on its applicability. \newline 

\begin{table*} [pos=h]

\caption{Strategies Used to Identify and Measure Technical Requirements Debt.}\label{Tab: RQ2}
\begin{tabular}{l c c c l}
\hline
\textbf{Strategy}                     & \textbf{Grounded} & \textbf{Density} & \multicolumn{1}{l}{\textbf{Amount}} & \multicolumn{1}{c}{\textbf{Primary Studies}}                                           \\ \hline
Manual management                     & 5                 & 4                & 9                                    & \begin{tabular}[c]{@{}l@{}}P72, P110, P125, P177, P196\end{tabular} \\ \hline
Automated management                  & 5                & 2                & 7                                    & \begin{tabular}[c]{@{}l@{}}P61, P125, P177, P196\end{tabular}        \\ \hline
Customer review                       & 4                 & 2                & 6                                    & \begin{tabular}[c]{@{}l@{}}P117, P125, P126, P178\end{tabular}       \\ \hline

Tools and softwares                   & 3                 & 2                & 5                                   & P61, P77, P177                                                                \\ \hline
Merge manual and automatic management & 2                 & 2                & 4                                    & P76, P134                                                                \\ \hline
Documentation template                & 3                 & 1                & 4                                    & P72, P79, P110                                                                \\ \hline
Face-to-face communication            & 2                 & 1                & 3                                    & P132, P178                                                               \\ \hline
Peer review                           & 2                 & 1                & 3                                    & P135, P178                                                               \\ \hline

Simple cost-benefit analysis          & 2                 & 0                & 2                                    & P23, P49                                                                 \\ \hline
Analytical hierarchy process          & 2                 & 0                & 2                                    & P13, P49                                                                 \\ \hline
Quantification approach               & 2                 & 0                & 2                                    & P15, P192                                                                \\ \hline
Approach of the nearest neighbor      & 2                 & 0                & 2                                    & P4, P99                                                                  \\ \hline
Cause and effect diagram              & 2                 & 0                & 2                                    & P183, P229                                                                      \\ \hline
Preventive actions              & 2                 & 0                & 2                                    & P203, P275                                                                      \\ \hline
RE-KOMBINE                         & 1                 & 1                & 2                                    & P25                                                                       \\ \hline

Backlog                               & 1                 & 0                & 1                                    & P76                                                                       \\ \hline
Payment map                              & 1                 & 0                & 1                                    & P208                                                                       \\ \hline
Identification through ISO/IEC/IEEE 29148:2018                             & 1                 & 0                & 1                                    & P263                                                                       \\ \hline
\end{tabular}
\end{table*}

\noindent \textbf{Manual management:} it refers to the process performed manually by software professionals. It would be to identify and measure the TD without the use of tools or software; \newline 

\noindent \textbf{Automated management:} in contrast to the management presented previously, the automated uses tools, software and automated resources to identify and measure the TD; \newline

\noindent \textbf{Customer review:} the work's success with quality requirements consists of involving stakeholders progressively, developing lists of sustainable requirements, and recording existing pending issues. Therefore, reviewing the requirements with the client, including the development team, should be considered essential for all management. Companies can adjust the product and specified requirements based on customer feedback to identify TDs more efficiently; \newline

\noindent \textbf{Tools and softwares:} the use of tools and software is present during automated management. When selecting one of these resources, the main question is related to the number of false positives that returns: how many TDs are analyzed in more detail and are not necessarily true. When these automated resources produce many false positives, this only distracts the location from the actual technical debts; \newline

\noindent \textbf{Merge manual and automatic management:} combining manual management with tool and software analysis is a practical and effective way to identify TDs in industrial projects; \newline

\noindent \textbf{Face-to-face communication:} in general, communication about requirements is hierarchical and based on e-mails and documents. But it is recommended that software professionals communicate directly with their peers and stakeholders. Face-to-face communication is efficient in the exchange of information between different stakeholders, helping to identify TDs. Therefore, face-to-face communication should be considered as essential in requirements management; \newline 

\noindent \textbf{Peer review:} benefits in reviewing requirements are highlighted in the literature, especially on defect identification and TDs. Among the forms of review, there is peer review. It consists of the analyst conducting the interview with the client and recording the audio of the dialogue. The audio is reviewed by another analyst (reviewer), who writes down ambiguities and lists the questions he would have asked if he had been the analyst. The questions are used for clarification in future interactions with the client; \newline

\noindent \textbf{Documentation template:} it refers to a model to be filled in to document TD, having different data that involve in particular its measurement. The TD documentation template proposed by \citet{seaman2011measuring} is shown in Table \ref{Tab: Template}.

\begin{table} [pos=ht]

\caption{Technical Debt Documentation Template.}\label{Tab: Template}
\begin{tabular}{{p{1.055in}p{1.92in}}}
\hline

\textbf{ID}                                   & Technical debt identification number                                                                 \\ \hline
\textbf{Date}                                 & Technical debt identification date                                                             \\ \hline
\textbf{Responsible}                                 & Person who identified the TD                                                             \\ \hline
\textbf{Location}                             & \begin{tabular}[c]{@{}l@{}}Description of where the debt item is \end{tabular}   \\ \hline
\textbf{Description}                                 & Justification of why that item needs to be considered                                                        \\ \hline
\textbf{Estimated
Principal}                           & Work required to pay off the TD                                                \\ \hline
\textbf{Estimated
Interest Amount}                           & Extra effort needed in the future if the TD item is not paid                                               \\ \hline
\textbf{Estimated
Interest
Probability}                        & \begin{tabular}[c]{@{}l@{}}Probability of extra work needed, if\\ the TD is not
paid off in the future\end{tabular} \\ \hline
\end{tabular}
\end{table}

\noindent \textbf{Simple cost-benefit analysis:} a list is created with TD items, where each one represents a task that has been left undone but is at risk of causing future problems.  It involves constructing a hierarchy of criteria (quantitative and qualitative criteria, objective and subjective, relevant for the decision), assigning weights and scales, and finally, making a series of comparisons between the items, indicating which should be paid first. Part of these criteria would be the definition of principal, interest and probability of interest. In the end, for example, a company may decide to approach 75 \% of the high interest bearing TDs, 25\% of the medium interest-bearing TDs and defer those with low interest. This strategy is also used to prioritize TD  \citep{lenarduzzi2021systematic}. \newline

\noindent \textbf{Analytical hierarchy process:} this process assigns weights and scales to different criteria used to measure the TD. Then, a series of pairwise comparisons are made between the items to obtain a prioritized TD classification. Based on this process, the items at the top of the list must be treated first. In other words, this strategy focuses on those TD items that have a potentially severe impact on the project regarding the total amount of interest that the project needs to pay. This strategy is not necessarily ideal, but it can decrease the project's risk level and keep the TD under control; \newline

\noindent \textbf{Quantification approach:} used to quantify TD and technical interest, but for this, it would be necessary to answer questions about your investments, such as: 1) How big is my TD? 2) How much interest am I paying for the TD? 3) Is the debt growing? and how fast? 4) What will be the consequence of keeping this TD for future maintenance? Finally, the technical requirements debt can be quantified considering the ratio between the user's needs that are already elicited and all possible user needs, including neglected ones; \newline

\noindent \textbf{Approach of the nearest neighbor:} this approach takes advantage of the experience acquired in the previously resolved TDs, based on the effort needed to correct these currently identified debts. The intuition is that the average time it takes to convert a TD of requirements is similar to the correction of previous debts in the project; \newline 

\noindent \textbf{Cause and effect diagram:} used to organize the causes that led to a TDs incidence, helping to identify it quickly, taking into account that the data are previously described; \newline 

\noindent \textbf{Preventive actions:} This strategy is not necessarily related to identifying and measuring the TD requirements, but it was thought essential to present it. Primary studies report that preventing the occurrence of a TD can be cheaper than its payment. Preventive actions support professionals and software teams in applying acceptable practices that minimize their occurrence. The preventive actions related to TD requirements are: controlling and negotiating the software requirements; well-defined requirement; good communication between stakeholders; well-defined scope statement; requirements change tracking, and customer commitment. \newline

\noindent \textbf{RE-KOMBINE:} it is a tool for the identification and management of TD requirements. This tool is based on the use of objective models used to and follow the software's evolution, identifying changes in requirements and understanding how they impact the current implementation of the software; \newline 

\noindent \textbf{Backlog:} often, part of the requirements documentation needs to be updated. However, there are more urgent tasks that need attention. In this case, write down the pending task in a TD list (similar to a daily task list), so that you don't lose sight of the correction that needs to be taken in the future. \newline 

\noindent \textbf{Payment map:} software professionals can consult this map to guide their decisions about eliminating TD in their projects. As a guide, the map can inform a set of practices and reasons in response to the need for TD payment. Furthermore, can be used as a communication device to support teams to effectively communicate existing TDs to managers to better decide on the payment of the TD requirements, for example. \newline

\noindent \textbf{Identification through ISO/IEC/IEEE 29148:2018:} this document references the quality of the requirements since specification to the other requirements engineering activities present in the software development life cycle. It provides details for the construction of consistent textual requirements, including characteristics and attributes and, language criteria of the requirements. From the moment that these quality standards are violated, requirements smells and technical debt may arise. Analyzing this document according to the project's specified and documented requirements helps to identify inconsistencies, low level of quality and violations, and, consequently, identify TD. \newline

\begin{figure}[ht]
	\centering
		\includegraphics[scale=.735]{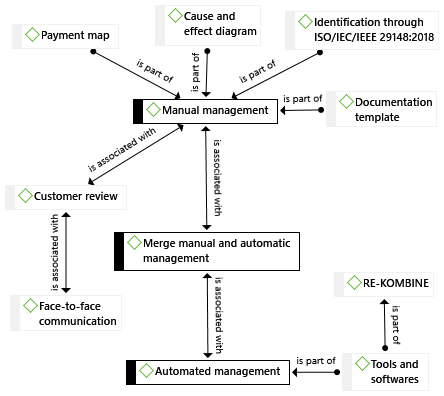}
	\caption{Association of Strategies for Identification and Measurement of the TD Requirements.}
	\label{FIG:9}
\end{figure}

Finally, among the 18 strategies, it was analyzed that 11 of them were related, as presented in Figure \ref{FIG:9}. It is possible to notice, for example, that the template of documentation of TD can be used during manual management. This type of management is associated with the review process with the client, which is strongly recommended to be performed in person. However, the manual management can also be related to the use of the payment map, the organization of the information and causes of a TD in the cause and effect diagram, and the analysis of the ISO/IEC/IEEE 29148:2018 standards. Automated management is associated with the use of tools and software, such as the RE-KOMBINE tool. Additionally, it is strongly recommended to use both management types to ensure an effective TD reduction process.

\subsection{RQ3: What metrics are being used to assist in the processof measuring the technical requirements debt?}

This question aimed to identify metrics that could help software professionals measure data related to the repayment of technical debt requirements. Initially, to provide a better understanding of the measurement stage of a TD, part of the primary studies presented definitions regarding the variables associated with technical debt that need to be calculated and measured, which are defined in detail in the sequence. \newline 

\noindent \textit{\textbf{ Principal: } } refers to the effort required to complete a task that has been left unattended. A task is a representation of a technical debt that is at risk of causing future problems if it is not repaid. The principal is calculated according to the number of technical debts that must be corrected in the software, the hours to fix each one, and the labor cost. \newline

\noindent \textit{\textbf{ Interest: }}
it is the penalty (in terms of more significant effort and lower productivity) that will have to be paid in the future due to the non-correction of technical debts at the present moment of identification. It refers to an estimate of the amount of extra work required to maintain the quality of the software if there is an unpaid technical debt. \newline

\noindent \textit{\textbf{ Interest Probability: }}
it refers to the likelihood that if the technical debt is not repaid, it will make other works more expensive over some time. The probability of interest is time-sensitive. \newline

In the sequence, were identified among the metric studies that help to measure: (i) the principal of TD; (ii) the interest on TD; (iii) the decision to reimburse TD at the time it was identified; (iv) the decision to reimburse TD at the time it was identified or at a specific point in the future; finally (v) the uncertainty about the measurement of TD. The purpose of this evidence is to assist in the decision on the reimbursement of TD requirements. Although it is not yet possible to present an accurate value in practice for each variable, these metrics are useful to reason and understand the factors involved during the measurement of technical debt, specifically about the possibility of adapting and using it in the context of software requirements. 

\subsubsection{Quantifying the Principal}

Following the recommendations of \citet{curtis2012estimating} (P258), the primary studies P33 and P177 presented a metric to calculate the Principal of the TD, being this a function of three main variables: (i) the number of TD items that should be reimbursed; (ii) the time needed to correct each item; and (iii) the cost to fix each TD item. The following metric are presented in sequence.

\begin{center} Principal = 
$$(  ( \sum high-severity \:TD ) \times ( percentage\: to\: be\: fixed )\times$$ 
$$( average\: hours\: needed\: to\: fix ) \times ( \$\: per\: hour ) ) + $$ 
$$(  ( \sum medium-severity\: TD  )\times  ( percentage\: to \: be\: fixed  )\times $$ 
$$ ( average\: hours\: needed\: to\: fix  ) \times ( \$\: per\: hour )  ) + $$
$$  (  ( \sum low-severity\: TD  ) \times ( percentage\: to\: be\: fixed  ) \times $$
$$ ( average\: hours\: needed\: to\: fix  )\times  ( \$\: per\: hour  ) ) $$

\end{center}

Following in the context of requirements, the number of TD items can be measured through a detailed analysis and review of the requirements specification documentation, for example. However, considering certain factors, such as the limited budget, software companies are rarely able to correct all the project's TDs. Therefore, each debt must be weighted by its severity, such as low, medium and high, to determine the percentage of TDs that will be reimbursed for each level.

Soon after, the time to correct a TD includes the time to analyze the debt, understand and determine its correction,  assess the potential side effects,  implement and test the correction, and the time to release the correction in operations \citep{curtis2012}. Finally, it is worth noting that this metric's variables can be adjusted to reflect better the company's experience and objectives, team or specific project.

\subsubsection{Quantifying the Interest}

As previously presented, TDs interest is the extra costs that will be spent on maintenance due to quality problems that will arise. In this sense, studies P15 and P43, state that the interest is the difference in maintenance effort between a certain level of quality and the ideal level. To estimate the Maintenance Effort (ME), the following metric is used:

\[ME = \frac{MF\times RV}{QF}\]

\noindent The metric above shows that the ME is a function of:

\noindent \textbf{Maintenance Fraction (MF): } represents the amount of maintenance effort that will be spent on an annual basis, measured as a percentage of changes involving updating requirements (added, modified or deleted) annually due to maintenance;

\noindent \textbf{Rebuild Value (RV):} is an estimate of the effort (person-months) that needs to be spent on rebuilding software, i.e. correct the existing TDs, determined by the metric:

\[RV = SS \times TF\]

System Size (SS) represents the total size of software measured in lines of code. Alternatively, SS can be measured using functional size (i.e. function points). Technology Factor (TF) represents the language's productivity factor, providing conversion to the effort (i.e., person-month). \newline

\noindent \textbf{Quality Factor (QF):} is a factor used to explain the level of quality of the software. It is assumed that the higher the quality level, the less effort is spent on maintenance. This statement is justified by previous research, which illustrates that making changes in software with superior quality is more efficient \citep{chidamber1998managerial}. The QF is determined from the following metric:

\[QF = 2^{  (   ( QualityLevel-3   ) \div 2  )}\]

According to primary studies, the above metric is a simplified model to consider the quality level (1 to 5 stars). For this purpose, the metric provides the following factors: 0.5, 0.7, 1.0, 1.4, 2.0, based on the work of  \citet{bijlsma2010indicators}.

\subsubsection{“If” Decision}

In study P96, a metric was presented to calculate whether it was worth paying back the TD when it was identified by relating the principal to interest. This metric came from the concept of \citet{seaman2012using}, which states that the TD should be paid if the principal is lower than the total interest.

\[\frac{CPrincipal}{TInterest} = result\]

CPrincipal is the current principal to be paid, and TInterest is the total interest in the software life cycle. The total interest usually cannot be calculated unless the software lifecycle is known, so TInterest is generally considered the interest calculated at a chosen point in the future. In other words, stakeholders need to understand if the repayment cost (principal) is lower than the total interest paid from now until the end of the software life cycle. From the analysis of the result, the following decisions can be considered:

\begin{itemize}
\item If the result is less than 1, it means that it is worth paying TD at present, i.e. the costs will be lower than those needed to refund in the future;
\item If the result is equal to 1, it means that there is no great loss in postponing the payment of TD, postponing the reimbursement will not accumulate in great costs; 
\item If the result is greater than 1, it means that it does not recommend paying TD at the moment it was identified, that is, the costs of reimbursement in the future will be lower.

\end{itemize}

Finally, the authors point out that choosing a point in the future before the final life cycle of the software is a safe choice. In fact, in the worst case, it is taken into account that the principal will cover only part of the interest. Furthermore, they point out that CPrincipal and TInterest are not described in terms of total costs in dollars, but in a set of factors associated with the process of reimbursement of TD requirements, for example. It is assumed that each element may be related to a positive or negative value that may increase or decrease over time.

\subsubsection{“When” Decision}

In primary study P96, a metric was presented to calculate the best time to repay the technical debt, whether in the current period in which it was identified or at a specific point in the future. For example, "should we reimburse now, or can we wait six months?" According to the authors, to answer this question, stakeholders need to know whether repaying at a chosen point in the future (F) is more or less convenient than repaying now. The metric follow in the sequence.

\[\frac{FPrincipal}{(TInterest - FInterest)} - \frac{CPrincipal}{TInterest} = result\]

The metric calculates the ratio between the principal in F and the remaining interest calculated as the total interest (TInterest) minus the interest paid in F. Soon after, it subtracts the proportion calculated about the same balance on the repayment in the current situation. From the analysis of the result, one can consider the following decisions:

\begin{itemize}
\item If the first term is greater than the second, the result will be a negative number, which means that it will be less convenient to pay TD later since the gain from repaying the debt about the remaining interest will be smaller than now;
\item If the result is low enough (close to 0), it means that repayment is not urgent, since it will not bring many benefits now concerning realizing it in the future. 
\end{itemize}

\subsubsection{Uncertainty of a Measurement }

According to \citet{curtis2012estimating}, there is no exact measurement for the TD, since the calculations are based only on structural failures that the organization needs to fix. In this sense, not all organizations reimburse the TDs based on appropriate techniques and metrics. Moreover, small changes in the variables related to a TD can cause large changes in its measurement, thus revealing final estimates' sensitivity.

Starting from this context, the primary study P43, states that calculations involving a TD can suffer systematic errors, caused, for example, by the low measurement that the tools present. With this, the authors define Uncertainty and Error, reporting that random errors or uncertainties in the measurement of a TD are frequent and refer to the delta that exists between the expected value of a measurement and its actual measurement. These errors can overestimate or underestimate the expected value of a measurement.

The primary study P43 used as a basis the operations and theories proposed by \citet{taylor1997introduction}, which states that the correct way to express the result of a measurement is to produce the best estimate of the greatness and the interval within which you are sure the greatness resides. In this sense, the authors of the primary study adapted to the context of TD, presenting the following metric:

\[measured \: value  \: of  \: TD_{principal} = (TD_{principal})_{best} \pm \delta_{TD} \] 

This metric represents the best estimate of a TD, with a margin of error or uncertainty about the principal TD \( ( \delta_{TD}) \). The estimate is between \( (TD_{principal})_{best} - \delta_{TD} \) \:  and \: \newline  \( (TD_{principal})_{best} + \delta_{TD} \). Following the recommendations of \citet{taylor1997introduction}, for convenience, uncertainty is always defined as positive, so that \( (TD_{principal})_{best} + \delta_{ TD} \) is always the most likely value of the greatness measured and \newline  \( (TD_{principal})_{best} - \delta_{ TD} \) is the least likely.

Finally, the authors of the primary study conclude that these calculations become appropriate in the measurement process, but that they still need to be validated in the domain of TD, mainly involving the different types of TD, for example, requirements. Furthermore, they reinforce the importance of understanding that the propagation of uncertainty is a critical factor that needs to be investigated to improve the decision-making process about which TD items to refactor.

\subsection{RQ4: What difficulties are pointed out during the management of technical debt requirements in software development?}

This question's objective was to identify the main difficulties pointed out by the studies in the process of managing the TD requirements. It is intended to present opportunities for new research, future work and development of new technologies, and assist software professionals in the difficulties reported, shown in the Table \ref{Tab: RQ4} in order of representativeness. Soon after, two networks were created for this issue in order to relate and associate the 18 difficulties identified.

\begin{table*} [pos=h]

\caption{Difficulties Identified in the Requirements Technical Debt Management Process.}\label{Tab: RQ4}

\begin{tabular}{l c c c l}
\hline
\textbf{Difficulty}                                                                                          & \textbf{Grounde} & \textbf{Densit} & \textbf{Amount} & \textbf{Pimary Studies}                                                     \\ \hline
Manage TD efficiently                                                                                        & 1                & 9               & 10              & P38                                                                         \\ \hline
Lack of access to tools                                                                                      & 6                & 1               & 7               & \begin{tabular}[c]{@{}l@{}}P76, P117, P154, P158, P178\end{tabular} \\ \hline
Understanding that TD is a problem                                                                           & 1                & 4               & 5               & P117                                                                        \\ \hline
Measure TD                                                                                                   & 3                & 1               & 4               & P4, P117, P146                                                              \\ \hline
Allocate more time in eliciting requirements                                                                 & 2                & 2               & 4               & P25, P178                                                                   \\ \hline
Engage the team in the TD management process                                                                 & 2                & 1               & 3               & P49, P117                                                                   \\ \hline
\begin{tabular}[c]{@{}l@{}}Balance the benefits of repaying TD with the costs\\ of this process\end{tabular} & 2                & 1               & 3               & P8, P132                                                                    \\ \hline
Manage unintentional TD                                                                                      & 2                & 1               & 3               & P9, P38                                                                     \\ \hline
Team morale                                                                                                  & 2                & 1               & 3               & P119, P228                                                                        \\ \hline
Adapt the team to the TD management process                                                                  & 1                & 2               & 3               & P111                                                                        \\ \hline
Automated management                                                                                         & 1                & 2               & 3               & P61                                                                         \\ \hline
Urgent management of TD                                                                                      & 1                & 2               & 3               & P24                                                                         \\ \hline
Customer collaboration in this process                                                                       & 1                & 1               & 2               & P74                                                                         \\ \hline
Culture or personal feelings                                                                                 & 1                & 1               & 2               & P50                                                                         \\ \hline
Manage older TDs                                                                                            & 1                & 1               & 2               & P38                                                                         \\ \hline
Conflicting goals                                                                                            & 1                & 1               & 2               & P50                                                                         \\ \hline
Organizational restrictions                                                                                  & 1                & 1               & 2               & P50                                                                         \\ \hline
Tradition                                                                                                    & 1                & 1               & 2               & P50                                                                         \\ \hline

\end{tabular}

\end{table*}

\begin{figure}[ht]
	\centering
		\includegraphics[scale=.8]{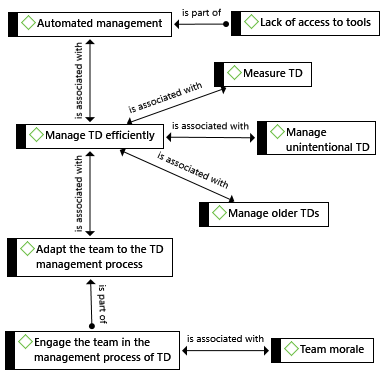}
	\caption{Difficulties Reported in Requirements TD Management (Network 1).}
	\label{FIG:10}
\end{figure}

The first network, illustrated in the Figure \ref{FIG:10}, presents nine difficulties reported during the management of TDs requirements. It becomes possible to notice, for example, that there is a difficulty in managing the TD efficiently, which causes an increase in maintenance costs at a rate that will eventually exceed the delivery value to the client.  This difficulty is associated with the TDs measurement because according to some reports, predicting the debt correction effort is often a more challenging task than predicting the effort to develop the software \citep{hassouna2010effort}.  

In the following, still, about managing the technical debt of requirements efficiently, primary studies report that unintentional TD is much more problematic to manage than intentional TD. Also, performing automated management along with the lack of tools in the context of requirements, become difficulties that compromise the quality of correction of the TD, because many projects do not have access to automated tools, besides the lack of sufficient infrastructure, which leaves teams hesitant to take on correction tasks. 

Finally, there is the difficulty in adapting the team to the management process of TD, because during the adaptation time the productivity usually drops, since the project or company has to go through the period of learning and education. This difficulty is associated with engaging the team in the management process because instead of only a few people documenting the TD requirements in the backlog, the collaboration of the other members would be necessary. This may be associated with team morale because if the TD is not corrected, it may hurt professionals' motivation. 

\begin{figure}[ht]
	\centering
		\includegraphics[scale=.74]{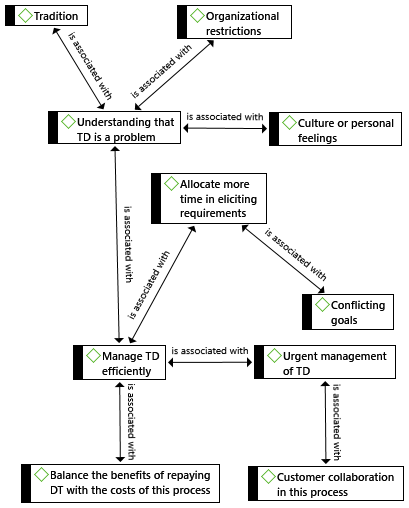}
	\caption{Difficulties Reported in Requirements TD Management (Network 2).}
	\label{FIG:11}
\end{figure}

In the second network created for this issue, which is presented in the Figure \ref{FIG:11}, it is possible to identify that the difficulty in managing the TD of requirements efficiently continues to relate to other codes, such as the urgency of the client's correction. This is associated with a lack of collaboration and communication that are essential elements in requirements engineering.

Soon after, there is the difficulty in balancing the benefits of managing the TD with the costs associated with this process, as well as the efforts spent on replanning the requirements, along with other implementation demands that the development teams have. Furthermore, there is difficulty allocating more time and effort to be paid during the elicitation of the requirements, often caused by conflicting goals, i.e., different objectives to be achieved in that period.

In the sequence, there is the need to understand that the technical debt of requirements if not managed correctly, can become a critical problem to the software project and be the root cause of other issues that arise. This difficulty can be associated with dealing with the team's culture or personal feelings, that is, people may prefer to always or never reduce a technical debt, this according to their senses and skills regardless of real interest. It may also be associated with organizational constraints, i.e., an action is only taken if the organization orders it.

Finally, there is the difficulty of tradition, that is, a particular practice is not changed or not included in the day to day of the project because it deviates from the standard way that the team performs the tasks.

\section{Discussion } \label{section5}

This systematic literature review aimed to investigate the current state of management research, specifically the identification and measurement of the TD requirements present in software development. To this end, 66 primary studies were analyzed to provide an overview of what has been discussed in the area by analyzing four research questions. 

This section presents a summary of the main discussions of the results that indicate their implications for software development industry professionals and researchers.

\subsection{Implications for Professionals}

The results have essential discussions for software development professionals, particularly those seeking guidance, strategies, tools and information in the literature that can help in certain situations they face in their projects.  The results of this SLR imply the following discussions for software development professionals:

\textit{1)} When 33 causes that cause the TD of requirements (RQ1) are presented, professionals are invited to a self-analysis about which of these inadequate actions exist in their projects. Furthermore, it assists in the identification of already existing TDs through these listed causes and in the verification of what can be improved to avoid the appearance of this debt;

\textit{2)} Among the causes mapped, there is a highlight to the low level of detail in the documentation of requirements, which is the root cause for the emergence of vague, incomplete and ambiguous requirements, for example. Through this analysis, it is recommended that industry and professionals dedicate more attention and effort to the process of specification and documentation of requirements, a more significant follow-up during the realization of these activities, along with a detailed review;

\textit{3)} It should be noted that part of the causes attributed to the emergence of the TD of requirements is associated with the client and stakeholders' collaboration. It is up to the industry professionals to tighten their communication during the life cycle of the requirement, ensuring its integrity according to what the client wants, avoiding incompatibilities and excessive updates throughout the software development;

\textit{4)} Most of the strategies that already help identify and measure the TD requirements (RQ2) are related to manual management. These are strategies that use documentation template, face-to-face communication, and payment map of TD, which becomes a challenge to implement in the industry, taking into account that most processes are automated due to agility and implementation time.  It is up to the professionals to analyze and adapt the strategies that best fit their needs, infrastructure and other related factors;

\textit{5)} Among the strategies identified, it is noticeable that two of them (Simple cost-benefit analysis and Analytical hierarchy process) can help professionals continue managing the TD requirements. Such strategies, besides helping in the measurement, with the assignment of weights and scales to the different criteria used in the estimates, can help prioritize the payment order of the identified items.  This happens when making comparisons between the TD items, concentrating the payment on those that have a profound potential impact on the project, for example;

\textit{6)} Knowing the causes for the emergence of TD requirements can support software development teams in defining actions that could have been taken to prevent these items of technical debt. Evidence in this context is presented in the strategy of preventive actions reported in RQ2. This information will help professionals use best practices that help prevent the emergence of this type of TD, and consequently avoid future costs and maintenance efforts.

\textit{7)} There is a lack of knowledge on the part of the industry on how to calculate the principal and mainly the interest of TD, which leads to increased costs and decreased quality of the software, which will possibly impact the clients. The metrics presented (RQ3) become a strategy that allows professionals to support these estimates, avoiding the accumulation of TD caused by inaccurate measurements;

\textit{8)} It is possible to notice that there is a lack of practical information in TD requirements. To solve this problem and, the software industry can benefit from these contributions, evaluations, tools and strategies coming from academia, it is necessary that software professionals play an active role in empirical studies, authorize, participate and provide the required data for academic research;

\textit{9)} Some of the difficulties reported (RQ4) are associated with measuring the TD of requirements, managing it efficiently and balancing the benefits with this process's costs. By analyzing this work, professionals will already find evidence to assist in solving these problems or provide support in conducting activities, through strategies and metrics. 

\subsection{Implications for Researchers}

The results present an active investigation area, but it still requires a deeper analysis, especially to validate the evidence presented in real cases. To guide future research, the results of this SLR imply the following discussions for researchers:

\textit{1)} There is an absence of tools and software that can be used in the context of TD requirements. Researchers are encouraged to develop such automated resources that support the process of managing this type of specific TD, especially on its measurement, adding the metrics identified to provide more accurate estimates;

\textit{2)} Empirical studies become necessary to present evidence in practice on applying the proposed metrics in the software industry. Part of the metrics are only mentioned in primary studies but not investigated in real cases. This research is necessary to provide refinement and adaptation of the metrics in the context of requirements so that they can be integrated into the working environment companies;

\textit{3)} Except RQ3, in the other questions networks were presented, which would be the relations between the codes (information) identified. According to the reports in studies, these networks aimed to associate principles that cause or influence others' existence. As a proposal for future research, it is recommended to confirm, in practice, whether these presented relationships are associated in real projects, or to what extent a cause for the emergence of the TD requirements can influence the existence of another cause, for example;

\textit{4)} Most investigations focus on technical aspects of the technical debt management process, but little attention is given to social and interpersonal issues. This is proven in RQ4 when part of the difficulties is related to professionals' culture and personal feelings, team morale, tradition, and organizational constraints. Research on these aspects becomes necessary to help overcome these problems and strengthen the process of managing TD requirements, proposing solutions that go beyond technical aspects;

\textit{5)} New research on identifying technical debt from the analysis of quality standards, norms and other information that an ISO provides becomes necessary, especially in the context of requirements. Analyze to what level software companies follow these quality standards, in which of them there is a more significant violation and, consequently, the appearance of TD, as well as the development of a strategy that facilitates the insertion and use of these ISO in the day to day of the companies. All these new researches will support the construction of consistent and quality requirements.

\textit{6)} Resistance on the part of companies in managing TD, engagement of the team in this process, and understanding that requirements debt can become a critical problem if not managed are among the difficulties reported. This proves the lack of research to strengthen collaboration between academia and industry. There is a need for training building so that professionals can understand the concept of TD, its consequences and differentiate it from other problems in their projects. From this understanding, facilitate the acceptance of companies in managing it, understanding the real benefits.

\textit{7)} Among all the strategies presented in RQ2, only one of them provides a payment map. That is, there is an absence of manuals, guides and mainly evidence briefings that gather the main information, evidence, strategies and tools on technical debt, either in a general context, or in its specific types, such` as requirements. The information presented in this standard would facilitate the acceptance and use of industry professionals. Considering the pressures of schedule and time to market, for example, providing information compiled, illustrated and quickly understood, would facilitate the insertion into the day-to-day of projects.

\section{Threats to Validity} \label{section6}

This section addresses the possible threats that affect this SLR results, and in the sequence, actions taken to minimize these biases are presented. To analyze these threats, the recommendations of  \citet{wohlin2012experimentation} were followed, which included
threats of construct validity, internal, external and conclusion, which are presented in sequence.

\textbf{\textit{Construct validity:}} this threat reflects what extent the operational measures that are studied represent what the researcher has in mind and what is investigated according to the research questions.  In this study, threats to construct reliability are considered: the decision of which studies to include or exclude and how to categorize them may have been biased and therefore, a threat. To minimize it, the selection processes were performed by at least three researchers. Additionally, the search string may not include all terms related to the search topic. However, pilots were previously performed to adjust the presented string.

\textbf{\textit{Internal validity:}} the threat to internal validity concerns the reliability of the methods used to analyze the data. The reliability of this work was high given the use of systematic and strict procedures and protocol, involvement and discussion with more than one researcher, and the use of a qualitative analysis tool for extraction and detailed analysis of data. 

\textbf{\textit{External validity:}} the threat to external validity refers to the generalization of results. In this context, this work focused on analyzing specific evidence to TD requirements, previously presenting conclusions only to this type of technical debt. Additionally, strategies and metrics for identification and measurement were introduced, which were analyzed conceptually but not validated in practice, not proving their generality to this or other TD types. It is concluded that maintaining a specific focus on research increases the threat of external validity. However, the results have allowed insights to guide new investigations.

\textbf{\textit{Conclusion validity:}} This threat relates to the validity of the conclusions drawn from the results. In this work, the non-identification and inclusion of primary studies are correlated, causing the loss of evidence related to the study's objective. Different data sources were first considered to include the maximum number of related studies to minimize the mentioned threat. Additionally, specialized and renowned scientific-academic sources and digital libraries were considered in computer science and, to the themes associated with the objective of this work. The search strategies were carefully applied, which returned a relevant quantity of articles, allowing to include as many results as possible.

\section{Conclusion} \label{section7}

When software development stages are pending or not properly executed, they cause a phenomenon known as TD, which if not managed at the time and in the correct way, leads to financial and operational losses.  TD can be present in the different phases of the software life cycle. However, specific types of this phenomenon still need to be investigated to assist in its management, for example, the TD of requirements. In this context, this work presents the results of a secondary study that aimed to answer the research question "How to assist in the identification and measurement of the technical requirements debt in software development?".

The research's focus was to identify the leading causes attributed to the emergence of technical requirements debt, existing strategies that help identify and measure it, metrics that can be used as a support during measurement, and difficulties reported in performing these activities. The process of conducting the systematic literature review used rigorous steps and detailed evidence analysis. 66 primary studies were included through manual and automatic searches, together with the snowballing method. These primary studies were returned from renowned research sources and digital bases, providing an overview and update of the last 10 years of state of the art (2010 to 2020).

The results of this systematic review show different contexts that can lead to the emergence of the TD of requirements, involving causes for intentional or unintentional TD, a collaboration of clients and stakeholders, elicitation and documentation of requirements and pressure of schedule, for example. Along with this evidence, other strategies that help identify and measure the TD of requirements were mapped, emphasising manual management, with different application proposals. However, the results highlight the lack of automated resources focused on this type of specific TD.

As support in the TD requirements measurement step, different metrics were identified. However, it still requires a validation and refinement in the context of requirements. Also, they already provide guidance and insights on the factors involved during the measurement of a TD, and its application and adaptation in specific contexts. Finally, it was found that the main difficulties in performing these activities are related precisely to the measurement stage, in addition to non-technical aspects such as team morale and engagement. 

In the end, it is concluded that the results present considerable evidence related to the objective of this work. Thus, this work has as main contributions: (i) the availability of information to help software professionals identify and measure the TD of requirements in their projects; (ii) broaden the understanding of topics related to TD and its management process, allowing identify how specific contents of this phenomenon can be fragmented; (iii) the availability of metrics that can be used in the measurement of a TD; (iv) by identifying the difficulties and gaps that are encountered when performing requirements TD management, become an opportunity for the development of new research; (v) presents evidence beyond code TD, for example, investigating another type of TD, exposing additional results in the area.

As future proposals, research will be invested based on the gaps identified in this work, adjusting with the evidence already placed and new empirical studies. Also, the development of a guide to support the TD requirements' identification and measurement is currently at early stage. The guide will consist of evidence from the literature and information gathered through survey in the software industry. Its objective is to help professionals identify and measure the existing TD requirements in their projects, knowing instructions and metrics to measure the data necessary for its resolution.

\section*{Acknowledgments}

The authors would like to thank Coordenação de Aperfeiçoamento de Pessoal de Nível Superior - Brazil (CAPES) -  Finance Code 001, for the financial support for the development of this research.

\appendix

\begin{appendices}

\section{Selected Primary Studies} 
{\small}
\begin{description}
\item [\textbf{[P4]}] Hassouna, A., \& Tahvildari, L. (2010). An effort prediction framework for software defect correction. Information and Software Technology, 52(2), 197-209.

\item [\textbf{[P8]}]Guo, Y., \& Seaman, C. (2011, May). A portfolio approach to technical debt management. In Proceedings of the 2nd Workshop on Managing Technical Debt (pp. 31-34).

\item [\textbf{[P9]}] Klinger, T., Tarr, P., Wagstrom, P., \& Williams, C. (2011, May). An enterprise perspective on technical debt. In Proceedings of the 2nd Workshop on managing technical debt (pp. 35-38).

\item [\textbf{[P11]}] Theodoropoulos, T., Hofberg, M., \& Kern, D. (2011, May). Technical debt from the stakeholder perspective. In Proceedings of the 2nd Workshop on Managing Technical Debt (pp. 43-46).

\item [\textbf{[P13]}] Seaman, C., \& Guo, Y. (2011). Measuring and monitoring technical debt. In Advances in Computers (Vol. 82, pp. 25-46). Elsevier.

\item [\textbf{[P15]}] Nugroho, A., Visser, J., \& Kuipers, T. (2011, May). An empirical model of technical debt and interest. In Proceedings of the 2nd workshop on managing technical debt (pp. 1-8).

\item [\textbf{[P23]}] Seaman, C., Guo, Y., Zazworka, N., Shull, F., Izurieta, C., Cai, Y., \& Vetrò, A. (2012, June). Using technical debt data in decision making: Potential decision approaches. In 2012 Third International Workshop on Managing Technical Debt (MTD) (pp. 45-48). IEEE.

\item [\textbf{[P24]}] Snipes, W., Robinson, B., Guo, Y., \& Seaman, C. (2012, June). Defining the decision factors for managing defects: A technical debt perspective. In 2012 Third International Workshop on Managing Technical Debt (MTD) (pp. 54-60). IEEE.

\item [\textbf{[P25]}] Ernst, N. A. (2012, June). On the role of requirements in understanding and managing technical debt. In 2012 Third International Workshop on Managing Technical Debt (MTD) (pp. 61-64). IEEE.

\item [\textbf{[P33]}] Curtis, B., Sappidi, J., \& Szynkarski, A. (2012, June). Estimating the size, cost, and types of technical debt. In 2012 Third International Workshop on Managing Technical Debt (MTD) (pp. 49-53). IEEE.

\item [\textbf{[P34]}] Letouzey, J. L. (2012, June). The SQALE method for evaluating technical debt. In 2012 Third International Workshop on Managing Technical Debt (MTD). IEEE.

\item [\textbf{[P38]}] Spínola, R. O., Vetrò, A., Zazworka, N., Seaman, C., \& Shull, F. (2013, May). Investigating technical debt folklore: Shedding some light on technical debt opinion. In 2013 4th International Workshop on Managing Technical Debt (MTD) (pp. 1-7). IEEE.

\item [\textbf{[P43]}] Izurieta, C., Griffith, I., Reimanis, D., \& Luhr, R. (2013, June). On the uncertainty of technical debt measurements. In 2013 International Conference on Information Science and Applications (ICISA) (pp. 1-4). IEEE.

\item [\textbf{[P45]}] Zazworka, N., Spínola, R. O., Vetro', A., Shull, F., \& Seaman, C. (2013, April). A case study on effectively identifying technical debt. In Proceedings of the 17th International Conference on Evaluation and Assessment in Software Engineering (pp. 42-47).

\item [\textbf{[P49]}] Codabux, Z., \& Williams, B. (2013, May). Managing technical debt: An industrial case study. In 2013 4th International Workshop on Managing Technical Debt (MTD) (pp. 8-15). IEEE.

\item [\textbf{[P50]}] Falessi, D., Shaw, M. A., Shull, F., Mullen, K., \& Keymind, M. S. (2013, May). Practical considerations, challenges, and requirements of tool-support for managing technical debt. In 2013 4th International Workshop on Managing Technical Debt (MTD) (pp. 16-19). IEEE.

\item [\textbf{[P61]}] Sneed, H. M. (2014, January). Dealing with Technical Debt in agile development projects. In International Conference on Software Quality (pp. 48-62). Springer, Cham.

\item [\textbf{[P72]}] Oliveira, F., Goldman, A., \& Santos, V. (2015, August). Managing technical debt in software projects using scrum: An action research. In 2015 Agile Conference (pp. 50-59). IEEE.

\item [\textbf{[P74]}] Soares, H. F., Alves, N. S., Mendes, T. S., Mendonça, M., \& Spinola, R. O. (2015, April). Investigating the link between user stories and documentation debt on software projects. In 2015 12th International Conference on Information Technology-New Generations (pp. 385-390). IEEE.

\item [\textbf{[P76]}] Suryanarayana, G., Samarthyam, G., \& Sharma, T. (2015). Repaying Technical Debt in Practice.

\item [\textbf{[P77]}] Tools for Repaying Technical Debt

\item [\textbf{[P79]}] Li, Z., Liang, P., \& Avgeriou, P. (2015, May). Architectural technical debt identification based on architecture decisions and change scenarios. In 2015 12th Working IEEE/IFIP Conference on Software Architecture (pp. 65-74). IEEE.

\item [\textbf{[P81]}] Abad, Z. S. H., \& Ruhe, G. (2015, August). Using real options to manage technical debt in requirements engineering. In 2015 IEEE 23rd International Requirements Engineering Conference (RE) (pp. 230-235). IEEE.

\item [\textbf{[P91]}] Mendes, T. S., de F. Farias, M. A., Mendonça, M., Soares, H. F., Kalinowski, M., \& Spínola, R. O. (2016, April). Impacts of agile requirements documentation debt on software projects: a retrospective study. In Proceedings of the 31st Annual ACM Symposium on Applied Computing (pp. 1290-1295).

\item [\textbf{[P96]}] Martini, A., \& Bosch, J. (2016, May). An empirically developed method to aid decisions on architectural technical debt refactoring: AnaConDebt. In 2016 IEEE/ACM 38th International Conference on Software Engineering Companion (ICSE-C) (pp. 31-40). IEEE.

\item [\textbf{[P99]}] Akbarinasaji, S., Bener, A. B., \& Erdem, A. (2016, May). Measuring the principal of defect debt. In Proceedings of the 5th International Workshop on Realizing Artificial Intelligence Synergies in Software Engineering (pp. 1-7).

\item [\textbf{[P101]}] Ghanbari, H. (2016, January). Seeking technical debt in critical software development projects: An exploratory field study. In 2016 49th Hawaii International Conference on System Sciences (HICSS) (pp. 5407-5416). IEEE.

\item [\textbf{[P110]}] Guo, Y., Spínola, R. O., \& Seaman, C. (2016). Exploring the costs of technical debt management–a case study. Empirical Software Engineering, 21(1), 159-182.

\item [\textbf{[P111]}] Yli-Huumo, J., Maglyas, A., \& Smolander, K. (2016). The Effects of Software Process Evolution to Technical Debt—Perceptions from Three Large Software Projects. In Managing Software Process Evolution (pp. 305-327). Springer, Cham.

\item [\textbf{[P117]}] Yli-Huumo, J., Maglyas, A., \& Smolander, K. (2016). How do software development teams manage technical debt?–An empirical study. Journal of Systems and Software, 120, 195-218.

\item [\textbf{[P119]}] Ghanbari, H., Besker, T., Martini, A., \& Bosch, J. Looking for Peace of Mind? Manage your (Technical) Debt.

\item [\textbf{[P125]}] Femmer, H., Fernández, D. M., Wagner, S., \& Eder, S. (2017). Rapid quality assurance with requirements smells. Journal of Systems and Software, 123, 190-213.

\item [\textbf{[P126]}] Mohagheghi, P., \& Aparicio, M. E. (2017). An industry experience report on managing product quality requirements in a large organization. Information and Software Technology, 88, 96-109.

\item [\textbf{[P132]}] Heikkilä, V. T., Paasivaara, M., Lasssenius, C., Damian, D., \& Engblom, C. (2017). Managing the requirements flow from strategy to release in large-scale agile development: a case study at Ericsson. Empirical Software Engineering, 22(6), 2892-2936.

\item [\textbf{[P133]}] Beer, A., Junker, M., Femmer, H., \& Felderer, M. (2017, September). Initial investigations on the influence of requirement smells on test-case design. In 2017 IEEE 25th International Requirements Engineering Conference Workshops (REW) (pp. 323-326). IEEE.

\item [\textbf{[P134]}] Femmer, H., Unterkalmsteiner, M., \& Gorschek, T. (2017, September). Which requirements artifact quality defects are automatically detectable? A case study. In 2017 IEEE 25th International Requirements Engineering Conference Workshops (REW) (pp. 400-406). IEEE.

\item [\textbf{[P135]}] Ferrari, A., Spoletini, P., Donati, B., Zowghi, D., \& Gnesi, S. (2017, September). Interview review: detecting latent ambiguities to improve the requirements elicitation process. In 2017 IEEE 25th International Requirements Engineering Conference (RE) (pp. 400-405). IEEE.

\item [\textbf{[P146]}] Ampatzoglou, A., Michailidis, A., Sarikyriakidis, C., Ampatzoglou, A., Chatzigeorgiou, A., \& Avgeriou, P. (2018, May). A framework for managing interest in technical debt: an industrial validation. In Proceedings of the 2018 International Conference on Technical Debt (pp. 115-124).

\item [\textbf{[P154]}] Charalampidou, S., Ampatzoglou, A., Chatzigeorgiou, A., \& Tsiridis, N. (2018, August). Integrating traceability within the ide to prevent requirements documentation debt. In 2018 44th Euromicro Conference on Software Engineering and Advanced Applications (SEAA) (pp. 421-428). IEEE.

\item [\textbf{[P158]}] Tsoukalas, D., Siavvas, M., Jankovic, M., Kehagias, D., Chatzigeorgiou, A., \& Tzovaras, D. (2018, September). Methods and Tools for TD Estimation and Forecasting: A State-of-the-art Survey. In 2018 International Conference on intelligent systems (IS) (pp. 698-705). IEEE.

\item [\textbf{[P161]}] Conejero, J. M., Rodríguez-Echeverría, R., Hernández, J., Clemente, P. J., Ortiz-Caraballo, C., Jurado, E., \& Sánchez -Figueroa, F. (2018). Early evaluation of technical debt impact on maintainability. Journal of Systems and Software, 142, 92-114.

\item [\textbf{[P165]}] de O. Passos, A. F., de Freitas Farias, M. A., de Mendonça Neto, M. G., \& Spínola, R. O. (2018, October). A Study on Identification of Documentation and Requirement Technical Debt through Code Comment Analysis. In Proceedings of the 17th Brazilian Symposium on Software Quality.

\item [\textbf{[P167]}] Bano, M., Zowghi, D., Ferrari, A., Spoletini, P., \& Donati, B. (2018, August). Learning from mistakes: An empirical study of elicitation interviews performed by novices. In 2018 IEEE 26th International Requirements Engineering Conference (RE) (pp. 182-193). IEEE.

\item [\textbf{[P171]}] Martini, A., Sikander, E., \& Madlani, N. (2018). A semi-automated framework for the identification and estimation of architectural technical debt: A comparative case-study on the modularization of a software component. Information and Software Technology, 93, 264-279.

\item [\textbf{[P175]}] Conejero, J. M., Rodríguez-Echeverría, R., Hernández, J., Clemente, P. J., Ortiz-Caraballo, C., Jurado, E., \& Sánchez-Figueroa, F. (2018). Early evaluation of technical debt impact on maintainability. Journal of Systems and Software, 142, 92-114.

\item [\textbf{[P177]}] Kouros, P., Chaikalis, T., Arvanitou, E. M., Chatzigeorgiou, A., Ampatzoglou, A., \& Amanatidis, T. (2019, April). Jcaliper: search-based technical debt management. In Proceedings of the 34th ACM/SIGAPP Symposium on applied computing (pp. 1721-1730).

\item [\textbf{[P178]}] Rindell, K., Bernsmed, K., \& Jaatun, M. G. (2019, August). Managing Security in Software: Or: How I Learned to Stop Worrying and Manage the Security Technical Debt. In Proceedings of the 14th International Conference on Availability, Reliability and Security (pp. 1-8).

\item [\textbf{[P183]}] Rios, N., Spínola, R. O., Mendonça, M., \& Seaman, C. (2019, May). Supporting analysis of technical debt causes and effects with cross-company probabilistic cause-effect diagrams. In 2019 IEEE/ACM International Conference on Technical Debt (TechDebt) (pp. 3-12). IEEE.

\item [\textbf{[P192]}] Lenarduzzi, V., \& Fucci, D. (2019, September). Towards a holistic definition of requirements debt. In 2019 ACM/IEEE International Symposium on Empirical Software Engineering and Measurement (ESEM) (pp. 1-5). IEEE.

\item [\textbf{[P196]}] Mendes, T. S., Gomes, F. G., Gonçalves, D. P., Mendonça, M. G., Novais, R. L., \& Spínola, R. O. (2019). VisminerTD: a tool for automatic identification and interactive monitoring of the evolution of technical debt items. Journal of the Brazilian Computer Society, 25(1), 1-28.

\item [\textbf{[P203]}] Freire, S., Rios, N., Mendonça, M., Falessi, D., Seaman, C., Izurieta, C., \& Spínola, R. O. (2020, March). Actions and impediments for technical debt prevention: results from a global family of industrial surveys. In Proceedings of the 35th Annual ACM Symposium on Applied Computing.

\item [\textbf{[P208]}] Freire, S., Rios, N., Gutierrez, B., Torres, D., Mendonça, M., Izurieta, C., ... \& Spínola, R. O. (2020). Surveying software practitioners on technical debt payment practices and reasons for not paying off debt items. In Proceedings of the Evaluation and Assessment in Software Engineering (pp. 210-219).

\item [\textbf{[P209]}] Behutiye, W., Seppänen, P., Rodríguez, P., \& Oivo, M. (2020). Documentation of quality requirements in agile software development. In Proceedings of the Evaluation and Assessment in Software Engineering (pp. 250-259).

\item [\textbf{[P214]}] Behutiye, W., Rodríguez, P., Oivo, M., Aaramaa, S., Partanen, J., \& Abhervé, A. (2020, August). How agile software development practitioners perceive the need for documenting quality requirements: a multiple case study. In 2020 46th Euromicro Conference on Software Engineering and Advanced Applications (SEAA) (pp. 93-100). IEEE.

\item [\textbf{[P216]}] Li, Y., Soliman, M., \& Avgeriou, P. (2020, August). Identification and Remediation of Self-Admitted Technical Debt in Issue Trackers. In 2020 46th Euromicro Conference on Software Engineering and Advanced Applications (SEAA) (pp. 495-503). IEEE.

\item [\textbf{[P225]}] Lenarduzzi, V., Fucci, D., \& Mendéz, D. (2020). On the perceived harmfulness of requirement smells: An empirical study. In Joint 26th International Conference on Requirements Engineering: Foundation for Software Quality Workshops, Doctoral Symposium, Live Studies Track, and Poster Track, Pisa; Italy, 24 March 2020 through 27 March 2020 (Vol. 2584).

\item [\textbf{[P228]}] Besker, T., Ghanbari, H., Martini, A., \& Bosch, J. (2020). The influence of Technical Debt on software developer morale. Journal of Systems and Software, 167, 110586.

\item [\textbf{[P229]}] Rios, N., Spínola, R. O., Mendonça, M., \& Seaman, C. (2020). The practitioners’ point of view on the concept of technical debt and its causes and consequences: a design for a global family of industrial surveys and its first results from Brazil. Empirical Software Engineering, 25(5).

\item [\textbf{[P233]}] de Freitas Farias, M. A., de Mendonça Neto, M. G., Kalinowski, M., \& Spínola, R. O. (2020). Identifying self-admitted technical debt through code comment analysis with a contextualized vocabulary. Information and Software Technology, 121, 106270.

\item [\textbf{[P247]}] Dar, H. S. (2020, August). Reducing Ambiguity in Requirements Elicitation via Gamification. In 2020 IEEE 28th International Requirements Engineering Conference (RE) (pp. 440-444). IEEE.

\item [\textbf{[P249]}] Panis, M. C. (2020, August). An Analysis of Requirements - Related Problems that Occurred in an organization Using a Mature Requirements Engineering Process. In 2020 IEEE 28th International Requirements Engineering Conference (RE) (pp. 291-299). IEEE.

\item [\textbf{[P258]}] Curtis, B., Sappidi, J., \& Szynkarski, A. (2012). Estimating the principal of an application's technical debt. IEEE software, 29(6), 34-42.

\item [\textbf{[P263]}] Femmer, H., Fernández, D. M., Juergens, E., Klose, M., Zimmer, I., \& Zimmer, J. (2014, June). Rapid requirements checks with requirements smells: Two case studies. In Proceedings of the 1st International Workshop on Rapid Continuous Software Engineering (pp. 10-19).

\item [\textbf{[P273]}] Robiolo, G., Scott, E., Matalonga, S., \& Felderer, M. (2019, November). Technical debt and waste in non-functional requirements documentation: An exploratory study. In International Conference on Product-Focused Software Process Improvement (pp. 220-235). Springer, Cham.

\item [\textbf{[P274]}] Lenarduzzi, V., Orava, T., Saarimäki, N., Systa, K., \& Taibi, D. (2019, September). An empirical study on technical debt in a finnish sme. In 2019 ACM/IEEE International Symposium on Empirical Software Engineering and Measurement (ESEM) (pp. 1-6). IEEE.

\item [\textbf{[P275]}] Rios, N., Mendes, L., Cerdeiral, C., Magalhães, A. P. F., Perez, B., Correal, D., ... \& Spínola, R. O. (2020, March). Hearing the voice of software practitioners on causes, effects, and practices to Deal with documentation debt. In International Working Conference on Requirements Engineering: Foundation for Software Quality. Springer, Cham.

\end{description}
\end{appendices}

%\input{Table/studies}
%\input{Table/studies2}

%\verb+\printcredits+ command is used after appendix sections to list 
%author credit taxonomy contribution roles tagged using \verb+\credit+ 
%in frontmatter.

\printcredits

\bibliographystyle{cas-model2-names}

\bibliography{paper.bib}

\vskip3pt

\bio{}
\textbf{Ana Melo} is a Master's student in the Post Graduate Program in Computer Engineering (PPGEC) at the University of Pernambuco (UPE), where she researches technical debt in software development. Bachelor's degree in Information Systems (2019) from the Federal Rural University of Pernambuco (UFRPE). Member of the REACT Research Labs Research Group.  His research areas of interest include software development, technical debt, and requirements engineering. Contact her accm@ecomp.poli.br. 

\endbio

\bio{}
\textbf{Roberta Fagundes} has a Post-Doctorate in Statistics (2015) from the Federal University of Pernambuco (UFPE), Brazil. She also holds a doctorate (2013) and a master’s degree (2006) in Computer Science from UFPE. Graduated in Telematics Technology (2002) from the Federal Center for Technological Education of Paraíba (CEFET-PB). She is currently an Adjunct Professor at the University of Pernambuco (2007) in Information Systems and Computer Engineering at the University of Pernambuco (UPE), Brazil. She is also a vice- coordinator and professor of the Graduate Program in Computer Engineering (PPGEC), where there are Masters and Doctorate courses. She is interested in researching in research in the area of Computer Science, with emphasis on Computational Intelligence. Contact her roberta.fagundes@upe.br. 	 

\endbio

\bio{}
\textbf{Valentina Lenarduzzi} is a postdoctoral researcher at the LUT University in Finland. Her primary research interest is related to data analysis in software engineering, software quality, software maintenance and evolution, with a special focus on Technical Debt. She obtained her Ph.D. in Computer Science at the Università degli Studi dell’Insubria, Italy, in 2015, working on data analysis in Software Engineering. She also spent 8 months as Visiting Researcher at the Technical University of Kaiserslautern and Fraunhofer Institute for Experimental Software Engineering (IESE) working on Empirical Software Engineering in Embedded Software and Agile projects. In 2011 she was one of the co-founders of Opensoftengineering s.r.l., a spinoff company of the Università degli Studi
dell’Insubria. Contact her valentina.lenarduzzi@lut.fi.   

\endbio

\bio{}
\textbf{Wylliams Santos} is adjunct professor at the University of Pernambuco (UPE), Brazil, where he leads the REACT Research Labs. Ph.D. in Computer Science (2018), Informatics Center (CIn) at Federal University of Pernambuco (UFPE), Brazil. MSc in Computer Science (2011), Informatics Center at Federal University of Pernambuco, Brazil. He undertook his sandwich Ph.D. (2015-2016) research at the Department of Com- puter Science and Information Systems (CSIS) of the University of Limerick, Ireland and in collaboration with Lero - the Irish Software Research Centre. His research areas of interest include management of software projects, agile software development, technical debt and industry-academia collaboration. Contact her wbs@upe.br.   

\endbio

%\bio{figs/pic1}
%Author biography with author photo.
%Author biography. Author biography. 

%\bio{figs/pic1}
%Author biography with author photo.
%Author biography. Author biography. Author biography.

\end{document}